\documentclass{article}

\usepackage[T1]{fontenc}
\usepackage[utf8]{inputenc}
\usepackage[margin=1in,top=1in,bottom=3cm]{geometry}
\usepackage{perpage}
\MakePerPage{footnote}
\usepackage{epsfig}
\usepackage{mathtools}
\usepackage{latexsym}
\usepackage{amsmath}
\usepackage{mathrsfs}
\usepackage{yfonts}
\usepackage{makeidx}
\usepackage{graphicx}
\usepackage{slashed}
\usepackage{hyperref}
\usepackage[titletoc]{appendix}

\hypersetup{citecolor=blue,colorlinks=true,linkcolor=blue,pdfstartview={FitH},linktoc=page}

\usepackage{chngcntr}
\usepackage{bm}
\usepackage{dsfont}
\usepackage{multicol}
\usepackage{nameref}
\usepackage{textcomp}
\usepackage{changepage}
\usepackage{lettrine}
\usepackage{enumitem}
\usepackage[sortcites,backend=bibtex,style=numeric,sorting=anyvt,doi=false,url=false,giveninits=true,isbn=false]{biblatex}
\AtEveryBibitem{\clearfield{month}}
\AtEveryBibitem{\clearfield{day}}
\usepackage{tikzsymbols}
\usepackage{amsthm,amssymb}
\usepackage{subcaption}
\usepackage{mathtools}\mathtoolsset{showonlyrefs=true}

\newtheorem{thm1}{Theorem}
\newtheorem{prop1}[thm1]{Proposition}
\newtheorem{lem1}[thm1]{Lemma}
\newtheorem{cor1}[thm1]{Corollary}
\newtheorem{note}[thm1]{Remark}

 \newtheoremstyle{TheoremNum}
        {\topsep}{\topsep}              
        {\itshape}                      
        {}                              
        {\bfseries}                     
        {.}                             
        { }                             
        {\thmname{#1}\thmnote{ \bfseries #3}}
    \theoremstyle{TheoremNum}

     \newtheoremstyle{TheoremNum}
        {\topsep}{\topsep}              
        {\itshape}                      
        {}                              
        {\bfseries}                     
        {.}                             
        { }                             
        {\thmname{#1}\thmnote{ \bfseries #3}\thmnumber{}}
    \theoremstyle{TheoremNum}

\renewcommand{\d}{\mathrm{d}}

\newcommand{\R}{\mathbb{R}}
\newcommand{\dl}{\partial}

\newcommand{\hf}{\frac{1}{2}}

\newcommand{\scrh}{{\mathscr H}}
\newcommand{\scrs}{{\mathscr S}}
\newcommand{\hlm}{{\scrh}^L_{r_{_-}}}
\newcommand{\hrm}{{\scrh}^R_{r_{_-}}}
\renewcommand{\hm}{{\scrh}_{r_{_-}}}

\newcommand{\hlp}{{\scrh}^L_{r_{_+}}}
\newcommand{\hrp}{{\scrh}^R_{r_{_+}}}              
\renewcommand{\H}{{\mathcal{H}}}
\makeindex

\title{On the Scattering of Waves inside Charged Spherically Symmetric Black Holes}
\author{Mokdad Mokdad\thanks{University of Burgundy --- France; email: remi.mokdad@u-bourgogne.fr }, Rajai Nasser\thanks{ETH Zürich --- Switzerland; email: rajai.nasser@inf.ethz.ch} }

\begin{document}

\maketitle
 \vspace{-1cm}
\begin{abstract}
	In this paper we show that there is a breakdown of scattering between the event horizon (or the Cauchy horizon) and an intermediate Cauchy hypersurface in the dynamic interior of a Reissner-Nordström-like black hole. More precisely, we show that the trace operators and their analytic counterparts, the inverse wave operators, do not have bounded inverses, even though these operators themselves are bounded. This result holds for the natural energy given by the energy-momentum tensor of the wave equation using the timelike vector field  of the Regge-Wheeler variable, which asymptotically becomes normal to the horizons. The behaviour of solutions at low spatial-frequencies and their behaviour at high angular momenta are the only obstructions causing this breakdown of scattering. The breakdown follows from an analysis of a $1+1$-dimensional wave equation with  exponentially decaying potential which we treat for general potentials, and we show that the breakdown is generic.   
	 
\end{abstract}

\vspace{0.5cm}


\vspace{0.1in}


\tableofcontents

\section*{Introduction}

\addcontentsline{toc}{section}{\protect\numberline{}Introduction}

Recently, there have been a growing interest in the study of test fields inside black holes, namely in the region bounded by the event horizon and the Cauchy horizon when the latter is present. This interest sprouts mainly from the connection to  Penrose's cosmic censorship conjecture and the expected instability of the Cauchy horizon due to perturbations caused by such fields when coupled to the gravitational field. These studies consist in part of scattering theories and energy estimates. Scattering theories have been extensively developed in the exterior regions over the years, which highlights the great importance of these theories. For reviews on this literature see \cite{mokdad_maxwell_2016,mokdad_conformal_2019,nicolas_conformal_2016,georgescu_asymptotic_2017,hafner_scattering_2004,daude_direct_2017,mihalis_dafermos_scattering_2018} and the references therein, see also Section \ref{sec:waveop}. However, until recently, few mathematical studies focused on scattering in the interior regions.

\begin{figure} 
	\centering
 	\includegraphics[width=0.8\linewidth]{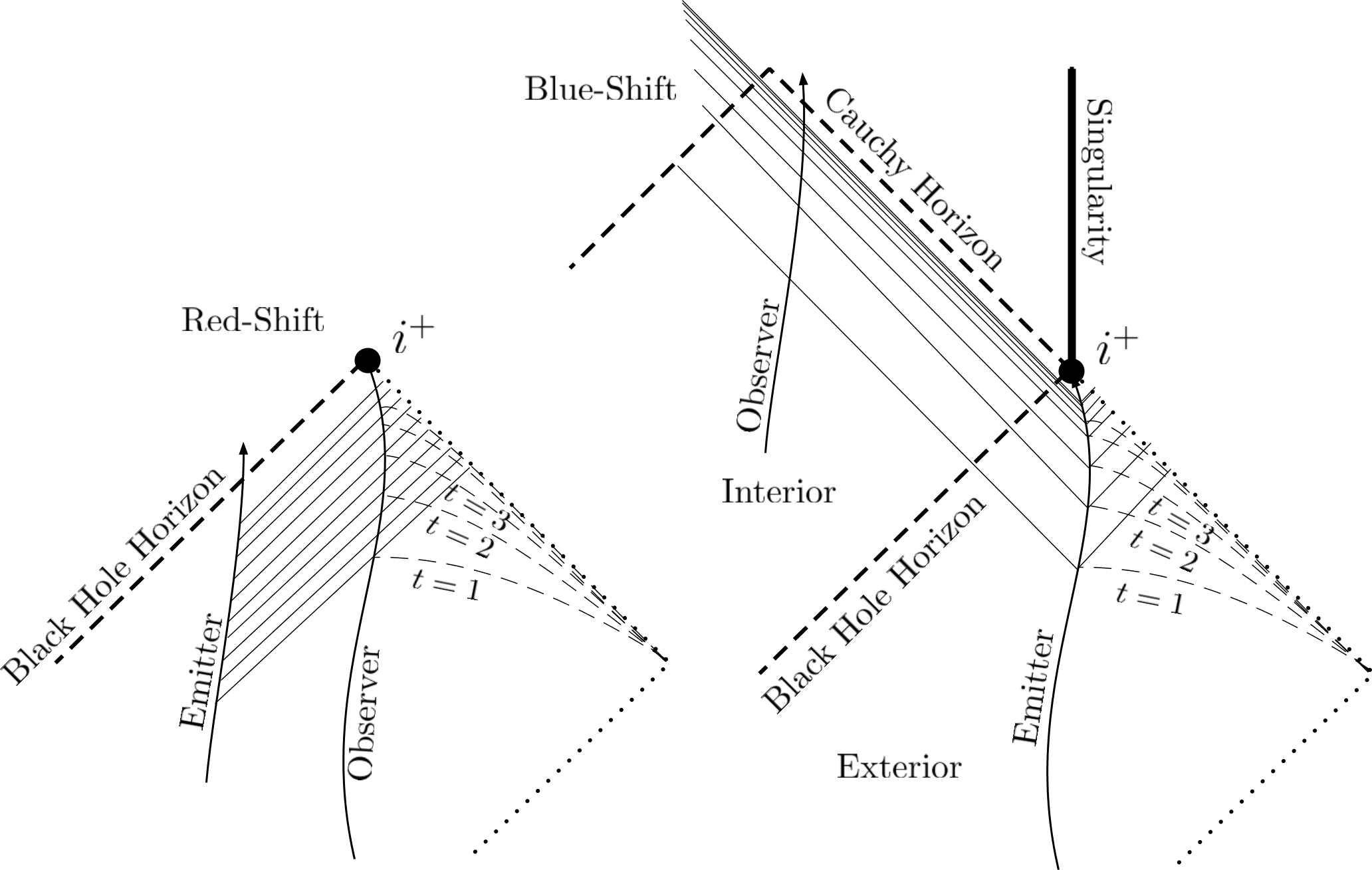}
	\caption{\emph{Illustration of gravitational red and blue shifts.}}\label{fig:BShift}
\end{figure}

R. Penrose and M. Simpson numerically observed \cite{simpson_internal_1973} that inside a Reissner-Nordström (RN) black hole, electromagnetic radiation scattering at the Cauchy horizon is blue shifted and the energy flux measured by an observer approaching the Cauchy horizon diverges (Fig.\ref{fig:BShift}). Motivated by these findings, S. Chandrasekhar and J.B. Hartle \cite{chandrasekhar_crossing_1982} investigated linearized gravitational perturbations of a RN black hole's interior using scattering and proved a similar blow-up in the flux of radiation near the Cauchy horizon. 
For  scattering of linear wave equations, the current main results were recently obtained by C. Kehle and Y. Shlapentokh-Rothman \cite{kehle_scattering_2019} in 2019. In the language of transmission and reflection coefficients, they constructed a complete scattering theory for the geometric wave equation in the interior of a RN black hole  from the event horizon \emph{directly} to the Cauchy horizon, and observed a breakdown of such scattering in the presence of a cosmological constant or a conformal mass for the Klein-Gordon equation. The fundamental cause of this scattering breakdown, its relation to the blue-shift, and its implications on the stability of the Cauchy horizon, are yet to be understood. On the other hand, using the approach of dynamic scattering that is based on comparison dynamics, in 2020, D. Häfner, M. Mokdad, and J.P. Nicolas obtained a complete scattering theory for charged and massive Dirac fields inside spherically symmetric black holes of Reissner-Nordström-(anti-)de-Sitter  type \cite{hafner_scattering_2021}. Their method uses the wave operators, which are geometrically interpreted afterwards  as the trace operators. Therefore, their construction is first done from a horizon to an intermediate spacelike Cauchy hypersurface and then to the other horizon, which have been the usual approach for the scattering in the exterior of black holes.  Shortly after, M. Mokdad  obtained a conformal scattering theory for Dirac fields in the same settings \cite{mokdad_conformal_2022} using the waves-reinterpretation method for  solving the Goursat problem and showing the surjectivity of the trace operators, thus bypassing the analytic scattering and directly obtaining the geometric result. 

Other recent results study  the boundedness of  waves in the interior of RN black holes in the context of the strong cosmic censorship conjecture. These results are also relevant for scattering. In particular, there are works by A. Franzen and J.L. Costa \cite{franzen_boundedness_2016,costa_bounded_2017} inside RN, and on the interior of Kerr backgrounds  \cite{franzen_boundedness_2020} also by A. Franzen. Such energy bounds and the techniques employed for them are useful for the construction of scattering as can be seen in the above mentioned scattering works. 

In \cite{kehle_scattering_2019}, the authors mentioned that, for scalar wave equations, they do not expect that the  intermediate scattering maps from the horizons to a spacelike hypersurface
to be bounded in the interior of RN black holes.\footnote{Thus, the boundedness of the scattering operator from the horizon to the horizon is not easily provable via the method of  ``concatenating'' intermediate operators, as is usually done in the exterior.} According to them, this may be due to gravitational blue shift effects. In this paper we show that indeed the trace operators do not have bounded inverses, despite being bounded themselves. 
The breakdown of scattering that we prove in this paper happens for large enough angular momentum and in the limit of zero spatial-frequency. More precisely, at high angular momenta, the behaviour of solutions to the geometric wave equation in the interior of a  Reissner-Nordström-(Anti-)de Sitter black hole is governed by a $1+1$-dimensional wave equation whose potential is positive, depends only on time, and goes exponentially to zero in the infinite times. This means that solutions with vanishing spatial derivatives have zero energy at  infinite past and infinite future, but infinite energy on any finite time-slice. Data of such solutions can be approached by finite energy data that are smooth and compactly supported. Intuitively, this is the main reason for unboundedness. The relation between our result and the blue shift effect needs more investigation. Regarding the blue shift, there is a very recent mathematical study by J. Sbierski \cite{sbierski_instability_2022} on the blue-shift instability at the sub-extremal Kerr Cauchy horizon for the linearised vacuum Einstein equations.

Our results hold for the natural energy given by the energy-momentum tensor of the wave equation using the spacelike slices of the Regge-Wheeler variable, and the associated vector field\footnote{This vector field is timelike in the interior of the black hole and asymptotically becomes normal to the horizons.}. We suspect that a similar breakdown of scattering happens for any generic timelike vector field that is asymptotically normal to the horizons, in the sense that either the trace operators or their inverses would be unbounded. Furthermore, we do not expect that using another spacelike foliation would help avoiding this breakdown.

This contrast between the behaviour of Dirac fields and other fields\footnote{E.g., scalar waves. We also expect electromagnetic fields to exhibit similar behaviours to scalar waves.}, regarding the boundedness of the intermediate scattering operators, is related to the existence of conserved norms. Dirac fields always possess a positive definite quantity which can be used as an $L^2$-norm and is conserved independently of the spacetime background --- the so-called Dirac's current (see e.g., \cite{hafner_scattering_2021}). For other fields, the conservation of the norm   may largely depend on the  geometry of the underlying spacetime. For example, in the dynamic interiors of RN black holes, solutions to the geometric wave equation do not have an associated conserved positive quantity.    
 
The present paper is organized as follows: In Section \ref{sec:GeomFrame}, we first introduce the geometric set-up modelling the black hole's interior. We then discuss the wave equation and its associated energies and function spaces. We also define the trace operators in this section,  prove their boundedness (Proposition \ref{prop:traceop}),  and state our main result (Theorem \ref{mainthm}) which asserts that the trace operators do \textit{not} have bounded inverses. Section \ref{sec:AnalyticFram} is devoted for the technical proof of the theorem and its analytic form. We start by defining an auxiliary quantity that is better suited for the analysis of the wave equation after rescaling the solution. This quantity is then used in the analytic framework that is first applied on a toy model (Section \ref{sec:toymodel}) before treating general potentials of exponential decay in Section \ref{sec:gen_case}, including that of the black hole. Most of the outcomes in this paper are based on a principle technical result (Proposition \ref{prop:EtozeroSequence-Vgeneral}) which proves the existence of a sequence of initial data with constant non-zero energy, but whose corresponding sequence of solutions for the wave equation has zero energy limit at infinite times. In Section \ref{sec:waveop}, we construct the inverse wave  operator and prove its boundedness (Theorem \ref{thm:Inv-wave-op}) in the general settings of Section \ref{sec:gen_case}. The unboudedness of its inverse, the wave operator, is also proved in Theorem \ref{thm:Inv-wave-op}. We then apply the same arguments to the black hole case and obtain similar results in Theorem \ref{thm:inverse-wave-blackhole}.

\paragraph{\large{Acknowledgements}} The authors would like to thank D. Häfner and J.-P. Nicolas for previous valuable discussions on the subject. On behalf of M. Mokdad, the IMB receives support from the EIPHI Graduate School (contract ANR-17-EURE-0002).

 \newpage
\section*{Notations and Conventions}

\addcontentsline{toc}{section}{\protect\numberline{}Notations and Conventions}

We summarize here some of the notations and conventions used in this paper.
\begin{itemize}
	\item  For two real functions $A$ and $B$, we write $A\lesssim B$ to indicate 
	$$\text{\guillemetleft there exists an absolute constant $C>0$ such that $A(x)\le C B(x)$ for all $x$\guillemetright.}$$ Furthermore, if we wish to emphasize the dependence of the (hidden) constant $C$ on a parameter $\ell$, we use the symbol $\lesssim_{\ell}$.
	
	\item  We say that $A$ and $B$ are  equivalent, and write $A\eqsim B$, if $A\lesssim B$ and $B \lesssim A$, i.e., there exist two constants $C_1>0$ and $C_2>0$ such that $C_1 A(x) \le B(x) \le C_2 A(x)$ for all $x$. As before, the dependence on a parameter $\ell$ is indicated by $\eqsim_{\ell}$.
	\item The set $\mathcal{C}_c^\infty(\mathcal{O})$ is the collection of smooth compactly supported functions on $\mathcal{O}$ with values in $\mathbb{C}$.
\end{itemize}

\section{The Geometric Framework}\label{sec:GeomFrame}

\subsection{Background Spacetime}

We study general spherically symmetric black holes of the Reissner-Nordström-(anti-)de-Sitter type. More specifically, we are interested in the dynamic interior of the black hole, which is enclosed by the Cauchy and the event horizons.

Let $0<r_-<r_+<+\infty$ and let $f:[r_-,r_+]\to\mathbb{R}$ be a smooth function satisfying the following properties:
\begin{itemize}
\item $f(r_\pm)=0$ and $f<0$ on $]r_-,r_+[$,
\item $f'(r_\pm)\neq 0$.
\end{itemize}

Let $(\mathcal{M},\mathbf{g})$ be the Lorentzian manifold $]r_-,r_+[_r\times\R_x\times\mathcal{S}_{\omega}^2$ endowed with the metric
\begin{equation}\label{RNDSmetric}
	\mathbf{g}=-\frac{1}{f(r)}\d r^2+f(r)\d x^2-r^2\d \omega^2 \; , \qquad   \mathrm{with} \quad \d \omega^2=\d \theta^2 + \sin(\theta)^2\d \varphi^2 \; .
\end{equation}

Note that when $f(r)=1-\frac{2M}{r}+\frac{Q^2}{r^2}-\frac{1}{3}\Lambda r^2$,  $(\mathcal{M},\mathbf{g})$ corresponds to the dynamic interior region of a Reissner-Nordström-(anti-)de-Sitter black hole with two inner horizons, where $M>0$ and $Q\neq 0$ are respectively the mass and the charge of the black hole, and $\Lambda$ is the cosmological constant. The specific conditions on the parameters $M,Q$, and $\Lambda$ of these black holes for having two inner horizons are $M>|Q|$ for $\Lambda=0$, while for cosmological cases ($\Lambda\neq 0$), the conditions can be found in \cite{mokdad_reissnernordstromsitter_2017, hafner_scattering_2021}.

We define the Regge-Wheeler coordinate $t$ by requiring that $$\frac{\d t}{\d r}=\frac1f\quad\text{and}\quad t(r_0)=0$$ for some arbitrarily fixed $r_0\in~]r_-,r_+[$. Note that $r$ is a smooth and strictly decreasing function of $t\in\R$. Furthermore, $\displaystyle\lim_{t\rightarrow \pm \infty} r(t)= r_\mp$ since $r_\pm$ are simple zeros of $f$.

In terms of the coordinates $(t,x,\omega)\in \R_t\times \R_x\times\mathcal{S}_{\omega}^2$, the metric takes the form

\begin{equation}\label{RNDSmetricINRstar}
	\mathbf{g}=-f(r)(\d t^2 - \d x^2)-r^2\d \omega^2. 
\end{equation}
The spacetime and time orientations are set so that $(t,x,\omega)$ is positively oriented and $\dl_t$ is future oriented. We denote the spacelike hypersurface at time $t$ as $\Sigma_t:=\{t=cst\}$, and its future-oriented unit normal  as $\eta_t :=(-f)^{-\hf}\dl_t$.

We would like to extend the spacetime $(\mathcal{M},\mathbf{g})$ to include the hypersurfaces at $r=r_\pm$, which will correspond to the black hole horizon at $r=r_+$ and the Cauchy horizon at $r=r_-$. In order to do that, we introduce the Eddington-Finkelstein variables $u = t-x$ and $v = t+x$. In the coordinate systems $(r,u,\omega)$ and $(r,v,\omega)$, the metric's expressions are:
\begin{align}
	\mathbf{g} &=  f(r) \d u^2 - 2 \d u \d r - r^2 \d \omega^2 \quad\text{and}\quad \mathbf{g} = f(r) \d v^2 - 2 \d v \d r - r^2 \d \omega^2 \, .
\end{align} 

\begin{figure} 
	\centering
	\includegraphics[scale=1.2]{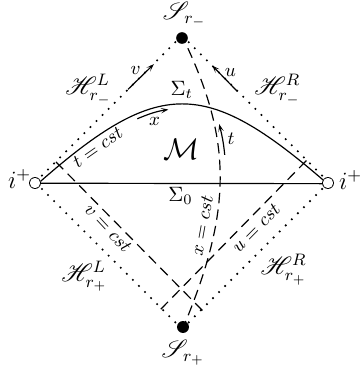}
	\caption{\emph{A Penrose-Carter conformal diagram  of $\bar{\mathcal{M}}$. 
	}}
	\label{fig:diagM}
\end{figure}

Therefore, the spacetime can be smoothly extended to $\mathcal{M}_u:=[r_-,r_+]_r\times \R_u  \times \mathcal{S}_{\omega}^2$ and to  $\mathcal{M}_v:=[r_-,r_+]_r\times \R_v\times \mathcal{S}_{\omega}^2$ which include the following four smooth null hypersurfaces

\begin{align*}
	\hlm &:= \{ r=r_-\}  \times \R_v \times \mathcal{S}_{\omega}^2 \subset \mathcal{M}_v \, ,\\
	\hrm &:= \{ r=r_-\}\times \R_u  \times  \mathcal{S}_{\omega}^2 \subset \mathcal{M}_u \, ,\\
	\hlp &:=  \{ r=r_+\} \times \R_u \times \mathcal{S}_{\omega}^2 \subset \mathcal{M}_u\, ,\\
	\hrp &:= \{ r=r_+\}\times \R_v  \times \mathcal{S}_{\omega}^2 \subset \mathcal{M}_v\, .
\end{align*}

The spacetime can be further extended to include the bifurcation spheres $\scrs_{r_\pm}$, which can be thought of as the place where ${\scrh}^L_{r_{_\pm}}$ and  ${\scrh}^R_{r_{_\pm}}$ meet. Refer to \cite{mokdad_reissnernordstromsitter_2017, kehle_scattering_2019} for an explicit construction of these spheres.

We refer to $\scrh_{r_-}=\hlm \cup {\scrs}_{r_{_-}} \cup \hrm$ as the Cauchy horizon and to $\scrh_{r_+}=\hlp \cup  {\scrs}_{r_{_+}}\cup \hrp$ as the event  horizon (or the black hole horizon). The quantities 
\begin{equation}
\kappa_-:=\hf f'(r_-)<0\quad\text{and}\quad\kappa_+:=\hf f'(r_+)>0
\end{equation}
are the surface gravities of the Cauchy and the event horizons, respectively. The extended spacetime is then $$\bar{\mathcal{M}}:=\mathcal{M}\cup\scrh_{r_+}\cup\scrh_{r_-}.$$ See Figure \ref{fig:diagM} for a Penrose diagram. 

The regions that are reached as $x\rightarrow\pm \infty$ along a curve of fixed $t$ (the regions labelled $i^+$ in Figure \ref{fig:diagM}) are not compactified. In the context of a sub-extremal Reissner-Nordström-(anti-)de-Sitter black hole where two distinct inner horizons exist, the $i^+$ region corresponds to future timelike infinity of the exterior region.

\subsection{Energies of the Wave Equation}

The geometric wave equation is given by 
\begin{equation}
\label{eq:geometricWaveEquation}
\square \phi=0,
\end{equation}
where locally $\square=\mathbf{g}^{ab}(\dl_a \dl_b  -\Gamma^c_{ab}\dl_c )$ with $\Gamma^c_{ab}$ being the Christoffel symbols of the metric $\mathbf{g}$. In the
%
%
 $(t,x,\theta,\varphi)$ coordinates, it has the form 
	\begin{equation}
	\square=\frac{1}{f}(\dl_x^2-\dl_t^2)-\frac{2}{r}\dl_t -\frac{1}{r^2} \Delta_{{\mathcal{S}^2}}. 
	\end{equation}

The geometric energies  associated to the wave equation are defined using its energy-momentum tensor 
\begin{equation}
	\mathbf{T}_{ab}:=\nabla_a \phi \nabla_b \phi -\frac{1}{2} \mathbf{g}_{ab}\nabla^c \phi \nabla_c \phi \; .
\end{equation}
It satisfies $\nabla^a \mathbf{T}_{ab}=\nabla_b \phi \square \phi$, which means that it is divergence-free (i.e., $\nabla^a \mathbf{T}_{ab}=0$) when $\phi$ is a solution for the wave equation. Additionally, the tensor $\mathbf{T}_{ab}$ satisfies the dominant energy condition\footnote{See e.g., \cite{hawking_large_1973}.}: For $X$ and $Y$ two causal vectors with the same time orientation\footnote{A causal vector is future-oriented if its inner product with the   vector field defining the time-orientation is positive.}, we have $\mathbf{T}_{ab}X^a Y^b\ge 0$.
One then defines the energy flux of the solution $\phi$ across a hypersurface $S$ with respect to a vector field $X$ to be:
\begin{equation}\label{enegrysurfacegeneralform}
\mathcal{E}_X [\phi](S):=\int_S \mathbf{T}_{ab}X^b \eta^a i_{\tau} \d  \mathrm{Vol}_{\mathbf g} ,
\end{equation}
for $\eta$ a normal vector to $S$ and $\tau$ a transverse one such that $\eta^a \tau_a=1$, with $\d \mathrm{Vol}_{\mathbf g}=-fr^2\d t\wedge\d x \wedge\d \omega^2$ the 4-volume form of the metric $\mathbf{g}$ and $i_{\tau} \d  \mathrm{Vol}_{\mathbf g}$ is the 3-volume form that is induced from $\d \mathrm{Vol}_{\mathbf g}$ on $S$ using $\tau$.  If $X$ is timelike and $S$ is spacelike, this quantity is non-negative. The definition is motivated mainly by the divergence theorem which in the case of a Killing vector field $X$ gives a conservation law. In the interior, however, there does not exist a timelike Killing  vector field, and therefore there is no obvious conserved positive quantity associated with the solution $\phi$.
\subsubsection{Energies on the Cauchy Hypersurfaces}
Here, $S$ will be one of our scattering surfaces $\Sigma_t$ which are spacelike with $\eta=\tau=\eta_t=(-f)^{-\hf}\dl_t$, and our timelike vector field $X$ will be $\dl_t$. Therefore, the geometric energy we use is positive and is given by
\begin{equation}\label{enegrygeometric}
\mathcal{E}[\phi](t):=\mathcal{E}_{\dl_{t}} [\phi](\Sigma_t)=\int_{\Sigma_t}\mathbf{T}_{00} r^2 \d x\wedge \d \omega^2=\hf\int_{\R_x\times\{t\}\times\mathcal{S}_{\omega}^2} \left((\dl_t \phi)^2 + (\dl_x \phi)^2 - \frac{f}{r^2}\vert \nabla_{\mathcal{S}^2}\phi\vert^2 \right)r^2 \d x \d^2\omega, 
\end{equation}
where $\displaystyle\nabla_{\mathcal{S}^2}\phi=\dl_{\theta}\phi\dl_\theta+\frac{\dl_{\varphi}\phi}{\sin(\theta)^2}\dl_\varphi$, $\vert \nabla_{\mathcal{S}^2}\phi\vert^2:=\d \omega^2(\nabla_{\mathcal{S}^2}\phi,\nabla_{\mathcal{S}^2}\phi)$, and $\d^2 \omega$ is the Lebesgue measure on $\mathcal{S}^2$.

We will also be interested  in the decomposition of the energy on the spherical harmonics. Set $$\phi_{\ell m} := \langle Y_{\ell m}, \phi\rangle_{L^2(\mathcal{S}^2)},$$ 
where $Y_{\ell m}(\theta,\varphi):=N P_{\ell m}(cos(\theta)) e^{im\varphi}$ are the spherical harmonics on $\mathcal{S}^2$, with $P_{\ell m}(cos(\theta))$ the associated Legendre polynomials, and $N$ is a normalization constant. Recall that $\Vert Y_{\ell m} \Vert_{L^2(\mathcal{S}^2)}=1$ and that $\Delta_{\mathcal{S}^2}Y_{\ell m}=-\ell(\ell+1)Y_{\ell m}$.  Note also that since $m$ appears nowhere in the equations (except through $\phi_{\ell m}$ itself), it plays no real role in our analysis. Thus, it is safe to drop it from the notation and set 
\begin{equation}\label{l=lm}
	\phi_\ell:=\phi_{\ell m}.
\end{equation}

We thus define
\begin{equation}\label{enegrygeometricharmonic}
\mathcal{E}_{\ell}[\phi](t)=\hf\int_{\R_x} \left((\dl_t \phi_{\ell})^2 + (\dl_x \phi_{\ell})^2 - \frac{f}{r^2}\ell(\ell+1)\phi_{\ell}^2 \right)r^2 \d x, 
\end{equation}
which can be obtained from $\mathcal{E}[\phi](t)$ after an integration by parts on the sphere and then decomposing on the harmonics.


\subsubsection{Energies on the Horizons}

The vector field $T:=\dl_{t}$ becomes $\dl_u+f\dl_r$ in the coordinate system $(r,u,\omega)$, and $\dl_v+f\dl_r$ in $(r,v,\omega)$, and therefore extends smoothly as $\dl_u$ to the horizons $\hrm$ and $\hlp$, and as $\dl_v$ to the horizons $\hrp$ and $\hlm$. $T$ is then normal to these horizons, while its inner product with the transverse vector field $N:=-\dl_r$ is unit, i.e., $T^aN_a=1$. Applying definition \eqref{enegrysurfacegeneralform}, we set
\begin{align}\label{enegryonhirzonsHrm&HLp}
	\mathcal{E}_T [\phi](\hrm)&=\int_{\hrm} \mathbf{T}_{ab}T^b T^a i_{N} \d  \mathrm{Vol}_{\mathbf g}=\int_{\hrm}{\mathbf{T}_ {00}r^2\d u  \wedge \d \omega^2} =\int_{\R_{u} \times \{r_-\} \times \mathcal{S}^2}{(\dl_{u} \phi)^2 r_-^2\d u \d^2 \omega}  ,\\
	\mathcal{E}_T [\phi](\hlm)&=\int_{\hlm} \mathbf{T}_{ab}T^b T^a i_{N} \d  \mathrm{Vol}_{\mathbf g}=\int_{\hlm}{\mathbf{T}_ {00}r^2\d v  \wedge \d \omega^2} =\int_{\R_{v} \times \{r_-\} \times \mathcal{S}^2}{(\dl_{v} \phi)^2 r_-^2\d v \d^2 \omega}  ,
\end{align}
and the energies on the past horizons at $r_+$ have similar expressions. In these notations, the energies on the total horizons are
\begin{equation}
	\mathcal{E}_T [\phi](\scrh_{r_\pm}) =	\mathcal{E}_T [\phi](\scrh_{r_\pm}^L)+	\mathcal{E}_T [\phi](\scrh_{r_\pm}^R).
\end{equation}

 We also decompose on the spherical harmonics to get
\begin{align}\label{enegryonhirzonsHrm&HLpONEHarmonic}
	\mathcal{E}_{T,\ell} [\phi](\hrm)&=\int_{\R_{u}}{(\dl_{u} \phi_{\ell})^2 r_-^2\d u }  ,\\
	\mathcal{E}_{T,\ell} [\phi](\hlm)&=\int_{\R_{v}}{(\dl_{v} \phi_{\ell})^2 r_-^2\d v }  ,\\
\mathcal{E}_{T,\ell} [\phi](\scrh_{r_\pm}) &=	\mathcal{E}_{T,\ell} [\phi](\scrh_{r_\pm}^L)+	\mathcal{E}_{T,\ell} [\phi](\scrh_{r_\pm}^R).
\end{align}

The following lemma relates the energy at the horizons with energies at Cauchy hypersurfaces.

\begin{figure} 
	\centering	
	
	\includegraphics[scale=1]{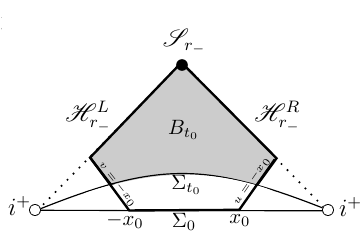}
	\caption{Bulk region for the divergence theorem.}
	\label{fig:divergence}
\end{figure}

\begin{lem1}
\label{lem:Divergence}
For every $\phi\in \mathcal{C}^\infty( \bar{\mathcal{M}})$ solution to Equation \eqref{eq:geometricWaveEquation} such that $\phi|_{\Sigma_0}$ and $\dl_t \phi|_{\Sigma_0}$ are compactly supported, we have
$$\mathcal{E}_{T} [\phi](\scrh_{r_\pm}) = \lim_{t\to\mp \infty}\mathcal{E}[\phi](t).$$
\end{lem1}
\begin{proof}
It is worth mentioning that if  $\phi|_{\Sigma_0}$ and $\dl_t \phi|_{\Sigma_0}$ are compactly supported, then since the propagation speed is 1, $\phi|_{\Sigma_{t_0}}$ is compactly supported for every $t_0\in\mathbb{R}$.

Let $x_0>0$ be such that $$\text{supp}\big(\phi|_{\Sigma_0}\big)\subset \{0\}_t\times]-x_0,x_0[_x\times\mathcal{S}^2_{\omega}.$$
For every $t_0\geq 0$, let $B_{t_0}$ be the bulk region of $\bar{\mathcal{M}}$ that is enclosed by the hypersurfaces $\Sigma_{t_0}$, $\hm$, $u=-x_0$ and $v=-x_0$ (see the shaded area in Figure \ref{fig:divergence}). If we apply the divergence theorem to $B_{t_0}$, we get
\begin{align*}
\int_{B_{t_0}} \nabla^a T^b \mathbf{T}_ {ab}\d \mathrm{Vol}_{\mathbf g}= \mathcal{E}_{T} [\phi](\scrh_{r_-}) - \mathcal{E} [\phi]({t_0}).
\end{align*}
Now
\begin{align*}
\int_{B_{t_0}} \nabla^a T^b \mathbf{T}_ {ab}\d \mathrm{Vol}_{\mathbf g}&\leq \left(\sup_{ B_{t_0}} \left| \nabla^a T^b \mathbf{T}_ {ab}\right|\right)\cdot \mathrm{Vol}_{\mathbf g}( B_{t_0})\\
&\leq \left(\sup_{B_0} \left| \nabla^a T^b \mathbf{T}_ {ab}\right|\right)\cdot \mathrm{Vol}_{\mathbf g}( B_{t_0}).
 \end{align*}
On the one hand, $\nabla^a T^b \mathbf{T}_ {ab}$ is smooth on $B_0\subset \bar{\mathcal{M}}$ and $B_0$ is compact, and thus,
 $$\sup_{B_0} \left| \nabla^a T^b \mathbf{T}_ {ab}\right|<+\infty.$$ 
On the other hand,
\begin{align*}
\mathrm{Vol}_{\mathbf g}( B_{t_0}) &= \int_{B_{t_0}}-fr^2\d t \wedge \d x  \wedge \d \omega^2= -\int_{t_0}^{+\infty} fr^2 \d t\int_{-x_0-t}^{x_0+t} \d x \int_{\mathcal{S}^2} \d \omega^2= -8\pi \int_{t_0}^{+\infty} (x_0+t)fr^2\d t.
\end{align*}
Now since $(x_0+t)fr^2\in L^1(\R_t)$ as it converges exponentially to 0 when $t\to+\infty$, it follows that
\begin{align*}
\mathrm{Vol}_{\mathbf g}( B_{t_0}) =-8\pi \int_{t_0}^{+\infty} (x_0+t)fr^2\d t\xrightarrow{t_0\to+\infty} 0.
\end{align*}
This shows that $\displaystyle \mathcal{E}_{T} [\phi](\scrh_{r_-}) = \lim_{t\to+ \infty}\mathcal{E}[\phi](t)$. The proof that  $\displaystyle \mathcal{E}_{T} [\phi](\scrh_{r_+}) = \lim_{t\to- \infty}\mathcal{E}[\phi](t)$ is similar.

%

\end{proof}

\subsubsection{Finite Energy Spaces}\label{sec:FiniteEnergySpaces}

Using the energy $\mathcal{E}_T$, one can define  Hilbert spaces of data with finite energy on the Cauchy slices and on the horizons. We define on  $\mathcal{C}_c^\infty(\Sigma_t)\times\mathcal{C}_c^\infty(\Sigma_t)$ the norm
\begin{equation}\label{energynorm-t}
\Vert( \phi_t, \psi_t) \Vert_{\mathcal{E}(t)}:=\left(\hf\int_{\Sigma_t}\left(\psi_t^2+(\dl_x\phi_t)^2  - \frac{f}{r^2}\vert \nabla_{\mathcal{S}^2}\phi_t\vert^2 \right)r^2 \d x \d^2\omega\right)^\hf,
\end{equation}
and we refer to it as the energy norm. Thus for a solution of $\square\phi=0$, $\Vert( \phi(t), \dl_{t}\phi(t)) \Vert_{\mathcal{E}(t)}^2:=\mathcal{E}[\phi](t)$. We define $\H(t)$, the finite energy space on $\Sigma_t$, to be the completion of $\mathcal{C}_c^\infty(\Sigma_t)\times\mathcal{C}_c^\infty(\Sigma_t)$ with respect to the energy norm \eqref{energynorm-t}. 

Similarly, we define the spaces $\mathcal{H}^\pm$ of functions with finite energies on the horizons. The finite energy space $\mathcal{H}^+$ on the Cauchy horizon $\scrh_{r_-}$ is the completion of $\mathcal{C}_c^\infty(\hlm)\times\mathcal{C}_c^\infty(\hrm)$ with respect to the norm
\begin{equation}
\label{eq:energyHplus}
	\Vert (\xi,\zeta)\Vert_{\H^+}=\left(\int_{\R_{u} \times \mathcal{S}^2}{(\dl_{u} \xi)^2 r_-^2\d u \d^2 \omega} +\int_{\R_{v} \times \mathcal{S}^2}{(\dl_{v} \zeta)^2 r_-^2\d v \d^2 \omega}\right)^\hf.
\end{equation}
$\H^-$ is defined analogously on $\scrh_{r_+}$. 

Let $\mathcal{C}^{\infty,0}_c(\scrh_{r_\pm})$ be the set of functions in $\mathcal{C}^{0}_c(\scrh_{r_\pm})$ -- i.e., that are continuous (in particular on bifurcation spheres) and compactly supported in the topology of $\scrh_{r_\pm}$ (i.e., supported away from the $i^+$ ``points'') -- but are also smooth on any point of the left and right parts $\scrh_{r_\pm}^L$ and $\scrh_{r_\pm}^R$ which are smooth 3-manifolds without boundaries, i.e.,

\begin{equation}
	\mathcal{C}^{\infty,0}_c(\scrh_{r_\pm})=\left\{\rho \in  \mathcal{C}^{0}_c(\scrh_{r_\pm}) : \rho|_{\scrh_{r_\pm}^L} \in \mathcal{C}^{\infty}(\scrh_{r_\pm}^L)   \text{ and } \rho|_{\scrh_{r_\pm}^R} \in \mathcal{C}^{\infty}(\scrh_{r_\pm}^R)  \right\}.
\end{equation}

In the next section, we will consider functions in $\mathcal{C}^{\infty,0}_c(\scrh_{r_\pm})$ that are also in $\mathcal{H}^\pm$. We therefore identify $$\rho \equiv \left(\rho|_{\scrh_{r_{\pm}}^L},\rho|_{\scrh_{r_{\pm}}^R}\right)\quad\text{for }\rho\in \mathcal{C}^{\infty,0}_c(\scrh_{r_\pm})\,.$$


\subsection{Trace Operators}



For every $(\phi_0,\psi_0)\in \mathcal{C}_c^\infty(\Sigma_0)\times\mathcal{C}_c^\infty(\Sigma_0)$, there exists unique $\phi\in \mathcal{C}^\infty( \bar{\mathcal{M}})$ solution to Equation \eqref{eq:geometricWaveEquation} satisfying the initial conditions $\phi|_{\Sigma_0} = \phi_0$ and $\dl_t\phi|_{\Sigma_0} = \psi_0$. To see this, consider the subset of $\bar{\mathcal{M}}$ consisting of the future and the past of the support of $(\phi_0,\psi_0)$, and note that this does not contain the ``points'' $i^+$. One can embed this subset in a globally hyperbolic spacetime extending it beyond the horizons, then using the classical theory for hyperbolic PDEs (e.g., Leray's theorems \cite{leray_hyperbolic_1955}), solve the Cauchy problem to obtain a solution that can then be restricted to the original subset. Furthermore, the restriction of $\phi$ to $\scrh_{r_\pm}$ is compactly supported. Accordingly, the future and past trace operators $\textgoth{T}^{\pm}:\mathcal{C}_c^\infty(\Sigma_0)\times\mathcal{C}_c^\infty(\Sigma_0) \to \mathcal{C}^{\infty,0}_c(\scrh_{r_\mp})$ are defined as $$\textgoth{T}^{\pm}(\phi_0,\psi_0) =\phi|_{\scrh_{r_{\mp}}} \equiv \left(\phi|_{\scrh_{r_{\mp}}^L},\phi|_{\scrh_{r_{\mp}}^R}\right)\,.$$

\begin{prop1}\label{prop:traceop}
We have 
\begin{equation}\label{CinftySubset}
\textgoth{T}^{\pm}\big(\mathcal{C}_c^\infty(\Sigma_0)\times\mathcal{C}_c^\infty(\Sigma_0) \big)\subseteq \H^{\pm}\,.
\end{equation}
Furthermore, the linear transformations $\textgoth{T}^{\pm}$ are bounded with respect to $\|\cdot\|_{\mathcal{E}(0)}$ and $\|\cdot\|_{\H^{\pm}}$. Therefore, $\textgoth{T}^{\pm}$ extend by continuity to linear bounded transformations $\textgoth{T}^{\pm}:\mathcal{H}(0)\to\H^{\pm}$, which we also refer to as the trace operators.
\end{prop1}
\begin{proof}
Let $(\phi_0,\psi_0)\in \mathcal{C}_c^\infty(\Sigma_0)\times\mathcal{C}_c^\infty(\Sigma_0)$, and let $\phi\in \mathcal{C}^\infty( \bar{\mathcal{M}})$ be the solution to Equation \eqref{eq:geometricWaveEquation} satisfying the initial conditions $\phi|_{\Sigma_0} = \phi_0$ and $\dl_t\phi|_{\Sigma_0} = \psi_0$. From Lemma \ref{lem:Divergence}, we have:
\begin{equation}\label{norm-of-trace=limit-energy}
\mathcal{E}_{T} [\phi](\scrh_{r_{\mp}})=\lim_{t\to\pm\infty }\mathcal{E}[\phi](t)\,.
\end{equation}

In order to obtain uniform bounds on the energy $\mathcal{E}[\phi](t)$, we first compute its derivative. From \eqref{enegrygeometric} we get
\begin{align}
\mathcal{E}[\phi]'(t) &= rf\int_{\Sigma_t} \left((\dl_t \phi)^2 + (\dl_x \phi)^2 - \frac{f}{r^2}\vert \nabla_{\mathcal{S}^2}\phi\vert^2 \right) \d x \d^2\omega\\
&\quad\quad+\frac{r^2}{2}\int_{\Sigma_t} \left(2\dl_t \phi\dl_t^2 \phi - 2\dl_t \phi\dl_{x}^2\phi + \frac{2f}{r^2}\dl_t\phi \Delta_{\mathcal{S}^2}\phi\right) \d x \d^2\omega-\hf\int_{\Sigma_t}  f\left(f'-\frac{2f}{r}\right) \vert \nabla_{\mathcal{S}^2}\phi\vert^2\d x \d^2\omega\\
&= \frac{2f}{r}\mathcal{E}[\phi](t) +\frac{r^2}{2} \int_{\Sigma_t} \left(-\frac{4f}{r}(\dl_t \phi)^2\right)\d x \d^2\omega -\hf\int_{\Sigma_t}  f\left(f'-\frac{2f}{r}\right) \vert \nabla_{\mathcal{S}^2}\phi\vert^2\d x \d^2\omega,
\label{eq:energyPhiDerivative}
\end{align}
where we applied integration by parts twice, once on $x$ and once on $\omega$, and we used the fact that $\phi$ satisfies \eqref{eq:geometricWaveEquation}. Note that
\begin{equation}
\label{eq:absoluteValDerivative}
\begin{aligned}
\left|\frac{2f}{r}\mathcal{E}[\phi](t)-\frac{r^2}{2} \int_{\Sigma_t} \left(-\frac{4f}{r}(\dl_t \phi)^2\right)\d x \d^2\omega\right| &\leq
\frac{6|f|}{r}\mathcal{E}[\phi](t).
\end{aligned}
\end{equation}
Therefore,
\begin{align}
|\mathcal{E}[\phi]'(t)| &\leq \frac{6|f|}{r}\mathcal{E}[\phi](t) +\left|f'-\frac{2f}{r}\right|\hf\int_{\Sigma_t} \left(\frac{-f}{r^2}\right)  \vert \nabla_{\mathcal{S}^2}\phi\vert^2 r^2\d x \d^2\omega\\
&\leq \frac{6|f|}{r}\mathcal{E}[\phi](t) +\left|f'-\frac{2f}{r}\right|\mathcal{E}[\phi](t). \label{eq:derivativeForLoose}
\end{align}
By Gr\"{o}nwall's lemma, we get for every $t_0\geq 0$

\begin{equation}
\label{eq:looseEstimate}
\frac{1}{C_{t_0}}\mathcal{E}[\phi](0)\leq\mathcal{E}[\phi](t_0)\leq C_{t_0} \mathcal{E}[\phi](0),
\end{equation}
where
\begin{equation}
\label{eq:defConstantCt0}
C_{t_0} = \exp\left(\int_0^{t_0}\left(\frac{6|f|}{r} +\left|f'-\frac{2f}{r}\right|\right)   \d t\right)<\infty.
\end{equation}
By applying \eqref{eq:absoluteValDerivative} to \eqref{eq:energyPhiDerivative}, we can also get
\begin{align*}
\mathcal{E}[\phi]'(t) &\leq \frac{6|f|}{r}\mathcal{E}[\phi](t) -\hf\int_{\Sigma_t}  f\left(f'-\frac{2f}{r}\right) \vert \nabla_{\mathcal{S}^2}\phi\vert^2\d x \d^2\omega.
\end{align*}
Now since $f'(r_-)<0$ and $f(r_-)=0$, it follows that there exists $t_0\geq 0$ such that $f'-\frac{2f}{r}<0$ for every $t\geq t_0$. Hence, $\mathcal{E}[\phi]'(t)\leq \frac{6|f|}{r}\mathcal{E}[\phi](t)$ for every $t\geq t_0$, and by Gr\"{o}nwall's lemma we get
$$\mathcal{E}[\phi](t)\leq  A_{t_0} \mathcal{E}[\phi](t_0),$$
where
$$A_{t_0} = \exp\left(\int_{t_0}^{+\infty}\frac{6|f|}{r} \d t\right)<\infty.$$
Since $C_t$ is an increasing function of $t$ and $A_{t_0}\geq 1$, we conclude that 
\begin{equation}\label{energy-estimates-t-positive}
\mathcal{E}[\phi](t) \leq A_{t_0}C_{t_0} \mathcal{E}[\phi](0), \quad \forall t\ge 0.
\end{equation}
Hence, from \eqref{norm-of-trace=limit-energy} and the definition of $\|\cdot\|_{\H^{+}}$, we have
\begin{equation}
\|\textgoth{T}^{+}(\phi_0,\psi_0)\|_{\H^{+}}^2=\mathcal{E}_{T} [\phi](\scrh_{r_{-}})=\lim_{t\to+\infty }\mathcal{E}[\phi](t) \leq A_{t_0}C_{t_0} \mathcal{E}[\phi](0)<\infty\,,
\end{equation}
i.e., $\textgoth{T}^{+}(\phi_0,\psi_0)\in\H^{+}$ proving \eqref{CinftySubset}, and
\begin{align*}
\|\textgoth{T}^{+}(\phi_0,\psi_0)\|_{\H^{+}}&\leq \sqrt{A_{t_0}C_{t_0} \mathcal{E}[\phi](0)}=\sqrt{A_{t_0}C_{t_0}} ~ \|(\phi_0,\psi_0)\|_{\mathcal{E}(0)}\,,
\end{align*}
which means that the future trace operator is bounded. 

For the past trace operator, we have from \eqref{eq:absoluteValDerivative} and \eqref{eq:energyPhiDerivative} that
\begin{align*}
\mathcal{E}[\phi]'(t) &\geq \frac{-6|f|}{r}\mathcal{E}[\phi](t) -\hf\int_{\Sigma_t}  f\left(f'-\frac{2f}{r}\right) \vert \nabla_{\mathcal{S}^2}\phi\vert^2\d x \d^2\omega.
\end{align*}
Let $\tilde{t}_0<0$ be such that $f'-\frac{2f}{r}>0$ for every $t\leq \tilde{t}_0$. A similar argument to the one above shows that 
\begin{equation}\label{energy-estimates-t-negative}
	\mathcal{E}[\phi](t) \leq \tilde{A}_{t_0}\tilde{C}_{t_0} \mathcal{E}[\phi](0), \quad \forall t\le 0,
\end{equation}
and that
\begin{align*}
\|\textgoth{T}^{-}(\phi_0,\psi_0)\|_{\H^{-}}&\leq\sqrt{\tilde{A}_{\tilde{t}_0}\tilde{C}_{\tilde{t}_0}} ~ \|(\phi_0,\psi_0)\|_{\mathcal{E}(0)},
\end{align*}
where
$$\tilde{C}_{\tilde{t}_0} = \exp\left(\int_{\tilde{t}_0}^{0}\left(\frac{6|f|}{r} +\left|f'-\frac{2f}{r}\right|\right)   \d t\right)<\infty\quad\text{and}\quad \tilde{A}_{\tilde{t}_0} = \exp\left(\int_{-\infty}^{\tilde{t}_0}\frac{6|f|}{r} \d t\right)<\infty.$$

\end{proof}

The main result of this paper is the following theorem:

\begin{thm1}\label{mainthm}
The extended trace operators $\textgoth{T}^{\pm}:\mathcal{H}(0)\to\H^{\pm}$ do not have bounded inverses.
\end{thm1}
\begin{proof}
The theorem follows immediately from Corollary \ref{cor:enegryEphiUnbounded} (\emph{cf}. Section \ref{sec:gen_case}) and identity \eqref{norm-of-trace=limit-energy}.
\end{proof}

\begin{note}
It is worth noting that one manifestation of the fact that the trace operator does not have a bounded inverse is the blow-up of the constant $C_{t_0}$ in the estimate \eqref{eq:looseEstimate} as $t_0\to\infty$. This happens because of non-integrability of the factor $\left|f'-2fr^{-1}\right|$ in  \eqref{eq:derivativeForLoose}. This factor comes from comparing the last term in \eqref{eq:energyPhiDerivative} with $\mathcal{E}[\phi](t)$, and it does not seem that a sharper estimate can be obtained for a general solution. However, if we restrict ourselves to solutions $\phi$ satisfying, for example,
\begin{equation}
\label{eq:condScattering}
\int_{\Sigma_t}  \vert \nabla_{\mathcal{S}^2}\phi\vert^2\d x \d^2\omega \leq D\int_{\Sigma_t}  \vert \dl_x\phi\vert^2\d x \d^2\omega,\quad \forall t\geq 0,
\end{equation}
for some absolute constant $D>0$, then we can obtain from \eqref{eq:energyPhiDerivative} and \eqref{eq:absoluteValDerivative} that
\begin{equation}
\begin{aligned}
|\mathcal{E}[\phi]'(t)| 
&\leq h(t)\mathcal{E}[\phi](t),
\end{aligned}
\end{equation}
where
$$h(t)=\frac{6|f|}{r}+\frac{D|f|}{2}\left|f'-\frac{2f}{r}\right|.$$
Now since $h$ is integrable, we obtain
\begin{equation}
\label{eq:tightEstimate}
\frac{1}{F}\mathcal{E}[\phi](0)\leq\mathcal{E}[\phi](t)\leq F \mathcal{E}[\phi](0), \quad \forall t\ge 0.
\end{equation}
where
$$F= \exp\left(\int_{0}^{+\infty}h(t)\d t\right)<\infty.$$

The trace operators become bounded bijections with bounded inverses from smooth compactly supported data satisfying \eqref{eq:condScattering} to its image, in view of \eqref{eq:tightEstimate}. Therefore they extend to Hilbert space isomorphisms from the completion of these data in $\mathcal{H}(0)$ to the completion of its images in $\mathcal{H}^\pm$. Note that these completions are \textbf{proper} Hilbert subspaces of  $\mathcal{H}(0)$ and $\mathcal{H}^\pm$, respectively. Hence, one can obtain scattering on such restricted initial data.  

One way to impose \eqref{eq:condScattering} is to restrict the initial data $(\phi|_{t=0},\dl_t\phi|_{t=0})$ to functions whose angular momentum $\ell$ is bounded from above, and whose spatial frequency (i.e., the dual variable in the Fourier transform with respect to $x$) is bounded away from zero. A similar construction of a (restricted) scattering theory can be preformed for these conditions on the frequency and the angular momentum. 

In fact, as we shall see, our proof for the breakdown of scattering (Theorem \ref{mainthm}) is a direct consequence of the behavior of solutions at high angular momenta and small spatial frequencies.
\end{note}

%
%
\section{The Analytic Framework} \label{sec:AnalyticFram}

%
%

This section is devoted to prove Corollary \ref{cor:enegryEphiUnbounded} that we used in the proof of Theorem \ref{mainthm}. The wave equation \eqref{eq:geometricWaveEquation} can be simplified by rescaling the solution and setting $u=r\phi$. Thus we have
\begin{equation}\label{wave-wavelikeeq}
rf\square \phi =\dl_x^2 u -\dl_{t}^2 u-  \frac{f}{r^2} \Delta_{{\mathcal{S}^2}}u+ \frac{ff'}{r} u \; ,
\end{equation}	
and since we work in the interior where $f<0$, it is enough to study
\begin{equation}\label{rescaledwaveeq}
\dl_{t}^2 u -\dl_x^2 u +  \frac{f}{r^2} \Delta_{{\mathcal{S}^2}}u- \frac{ff'}{r} u  =0 \; .
\end{equation}
Moreover, decomposing on spherical harmonics, \eqref{rescaledwaveeq} becomes:
\begin{equation}\label{wavedecompeq}
  \dl_t^2 u_\ell - \dl_{x}^2 u_\ell + V_\ell(t) u_\ell =0 \; ,
\end{equation}
where the potential
\begin{equation}\label{VpotentialBH}
 V_\ell(t)=-\frac{f\big(r(t)\big)}{r^2(t)}\Big( \ell(\ell+1)  + r(t)f'\big(r(t)\big)\Big)
\end{equation}
may change sign for some $t\in\R$ corresponding to some $r$ inside the interval  $]r_-,r_+[$. In particular, since $f(r_-)=f(r_+)=0$ and $f<0$ inside the interval, $f'$ must change sign inside it, hence $V_0$ must change sign in $]r_-,r_+[$. However, after some $\ell_0$ large enough, we have $V_\ell>0$ for all $\ell>\ell_0$. 


\subsection{Auxiliary Energy}

Because of the rescaling of the solution, $\mathcal{E}_{\ell}[\phi](t)$ is not the most convenient quantity to use when working with the form \eqref{wavedecompeq} of the wave equation. We therefore use the following auxiliary energy:
\begin{equation}\label{enegryEellu}
E_{\ell}[u](t)=\int_{\R_x} (\dl_t u_\ell)^2 + (\dl_x u_\ell)^2 + V_{\ell} u_\ell^2  \d x. 
\end{equation}

\begin{note}
We will mostly consider functions $\phi$ and $u$ that are supported on one harmonic mode $(\ell,m)$ , i.e., $\phi=\phi_\ell Y_{\ell m}$ and $u=u_\ell Y_{\ell m}$ (see \eqref{l=lm}). In this case, we have
$$\mathcal{E}[\phi](t) = \mathcal{E}_{\ell}[\phi](t) = \hf\int_{\R_x} \left((\dl_t \phi)^2 + (\dl_x \phi)^2 - \frac{f}{r^2}\ell(\ell+1)\phi^2 \right)r^2 \d x, $$
and
$$E_{\ell}[u](t) = \int_{\R_x} (\dl_t u)^2 + (\dl_x u)^2 + V_{\ell} u^2  \d x. $$
\end{note}

The following simple lemma shows that we lose no information relevant to our results by considering $E_{\ell}[u]$ instead of $\mathcal{E}_{\ell}[\phi]$.
\begin{lem1}
\label{lem:equivEphiEu}

Let $\ell_0>0$ be the smallest $\ell$ for which $\ell(\ell+1)+rf'>0$. For all $\ell\geq\ell_0$, we have $E_{\ell}[u]\eqsim\mathcal{E}_{\ell}[\phi]$, i.e., the energies $E_{\ell}[u]$ and $\mathcal{E}_{\ell}[\phi]$ are uniformly equivalent in $t\in\R$.
\end{lem1}
\begin{proof}
We may assume without loss of generality that $\phi=\phi_\ell$ and hence $u=u_\ell$. Recall that $u=r\phi$ and 
$$ V_\ell=-\frac{f}{r^2}\left( \ell(\ell+1)  + rf'\right),
$$
and so $V_\ell>0$ for $\ell\geq\ell_0$. We have
	\begin{align}
	E_{\ell}[u](t)&=\hf\int_{\R_x} (\dl_t (r\phi))^2 + r^2(\dl_x \phi)^2 +r^2 V_{\ell}\phi^2  \d x \\\
	&=\hf\int_{\R_x} r^2(\dl_t\phi)^2 + 2rf\phi \dl_t\phi + f^2\phi^2 + r^2(\dl_x \phi)^2 +r^2 V_{\ell}\phi^2  \d x \\\
	&\le\int_{\R_x} f^2\phi^2 + r^2(\dl_t \phi)^2 + r^2(\dl_x \phi)^2 +r^2 V_{\ell}\phi^2 \d x \\\
	&\lesssim \mathcal{E}_{\ell}[\phi](t) -\int_{\R_x} f(-f+f'r)\phi^2 \d x \ \\
	&\lesssim \mathcal{E}_{\ell}[\phi](t)  -\int_{\R_x} f\ell(\ell+1) \phi^2 \d x \ \lesssim \mathcal{E}_{\ell}[\phi](t)
	\end{align}
    where we have used the fact that $ -f+f'r\lesssim 1 \le \ell(\ell+1)$ since $-f+f'r$ is a bounded function of $r$ on the compact interval $[r_-,r_+]$.
	Similarly, 
	\begin{align}
     \mathcal{E}_{\ell}[\phi](t)&=\hf\int_{\R_x} r^2\left(\dl_{t}\left(\frac{u}{r}\right)\right)^2+(\dl_{x}u)^2-\frac{f}{r^2}\ell(\ell+1)u^2 \d x \\\
	 &=\hf\int_{\R_x} (\dl_{t}u)^2 - \frac{2fu\dl_t u}{r} + \frac{f^2u^2}{r^2} +(\dl_{x}u)^2-\frac{f}{r^2}\ell(\ell+1)u^2 \d x \\\
	  &\leq\int_{\R_x} (\dl_{t}u)^2 + \frac{f^2u^2}{r^2} +(\dl_{x}u)^2-\frac{f}{r^2}\ell(\ell+1)u^2 \d x \\\
     &\lesssim E_{\ell}[u](t)-\int_{\R_x} \frac{f}{r^2}(-f+\ell(\ell+1))u^2 \d x \lesssim E_{\ell}[u](t)
     \end{align}
	where the last inequality follows from $\displaystyle-\frac{f}{r^2}(-f+\ell(\ell+1))\lesssim V_\ell$ for $\ell\geq \ell_0$, because
\begin{align}
\frac{\displaystyle-\frac{f}{r^2}(-f+\ell(\ell+1))}{V_\ell} = \frac{-f+\ell(\ell+1)}{\ell(\ell+1) + rf'} &\leq  \frac{-f+\ell_0(\ell_0+1) + \ell(\ell+1) - \ell_0(\ell_0+1)}{\ell_0(\ell_0+1) + rf' + \ell(\ell+1) - \ell_0(\ell_0+1)} \\ & \leq \max\left\{  \frac{-f+\ell_0(\ell_0+1) }{\ell_0(\ell_0+1) + rf'}~ , 1 \right\}\lesssim 1,
\end{align}
where the last estimate follows from the boundedness of $-f+\ell_0(\ell_0+1)$ and the fact that $\ell_0(\ell_0+1) + rf'>0$ on the compact set $[r_-,r_+]$.
	
\end{proof}

\subsection{General Potential}

Our proofs do not depend on the exact form of the potential $V_{\ell}$ defined in \eqref{VpotentialBH} but rather on some of its properties which we summarize in the following lemma.
\begin{lem1}
\label{lem:properties_V_ell}
	Let $\ell_0$ be as in Lemma \ref{lem:equivEphiEu} and let $V_\ell$ be given as in \eqref{VpotentialBH}, then for $\ell\ge\ell_0$
	\begin{enumerate}[label=\roman*.]
		\item $V_\ell(t)>0$ on $\R$.
		\item $\dl_{t}V_{\ell}(t)<0$ for $ t >t_{large}\ge 0$ and $\dl_{t}V_{\ell}(t)>0$ for $t <-t_{large}\le 0$.
		\item $V_\ell\eqsim -f\eqsim e^{2\kappa_-t}$ for $t>t_{large}\ge0$ and $V_\ell\eqsim -f\eqsim e^{-2\kappa_+t}$ for $t<-t_{large}\le0$.
	\end{enumerate}
Here $t_{large}$ depends on the black hole's parameters, i.e., on $f$. 
\end{lem1}
\begin{proof}
 The first point as well as $V_\ell\eqsim -f$ are immediate since $f'$ is bounded. For $ii.$, we have
 \begin{equation}
 \dl_{t}V_{\ell}(t)= V_{\ell}\left(f'-2\frac{f}{r}\right)+\frac{f^2}{r^2}(f'+rf''), 
 \end{equation}
 thus
 \begin{equation}
 	 \lim\limits_{t\rightarrow\pm\infty}\frac{\dl_{t}V_{\ell}(t)}{-f}=\frac{2\kappa_\mp}{r_\mp^2}(\ell(\ell+1)+2\kappa_\mp r_\mp),
 \end{equation}
where $2\kappa_+=f'(r_+)>0$ and $2\kappa_-=f'(r_-)<0$.
 Next, since $\dl_{t} f=f' f$, we have
 \begin{align}
 f(r(t))&=f(r(0))e^{\int_{0}^{t}f'(r(s))ds}\\
 &=f(r(0))e^{\int_{0}^{t}f'(r(s))-f'(r_-)ds+tf'(r_-)}\\
 &=f(r(0))e^{2\kappa_- t+w(t)}
 \end{align}
 where $w(t)=\int_{0}^{t}f'(r(s))-f'(r_-)ds$. Clearly, $w$ is  smooth since $r$  and $f$ are smooth, and it is bounded because $f'(r(s))-f'(r_-)\in L^1(\R^+)$. Indeed,
 \begin{equation}
 \int_{0}^{+\infty}f'(r(s))-f'(r_-)ds=\int_{r(0)}^{r_-}\frac{f'(r)-f'(r_-)}{f(r)}dr
 \end{equation}
 is finite as the only zero of $f$ in $[r_-,r(0)]$ is $r_-$ and it is a simple zero, so 
 \begin{equation}
\lim\limits_{r\rightarrow r_-}\frac{f'(r)-f'(r_-)}{f(r)}=\frac{f''(r_-)}{f'(r_-)}\in\R.
 \end{equation}
Therefore, there exist $w_0$ and $w_1$ in $\R$ such that $w_0\le w(t)\le w_1$ for all $t\ge 0$, from which it follows that $-f\eqsim e^{2\kappa_-t}$ for $t>t_{large}\ge0$. The behaviour of $f$ near $r_+$ can be obtained analogously. 
\end{proof}

Hence, in this section we treat a general form of Equation \eqref{wavedecompeq}:
\begin{equation}\label{waveAnalytic}
\dl_t^2 u(t,x) - \dl_{x}^2 u(t,x) + V(t) u(t,x) =0 \; ,  ~~ (t,x)\in\R_t^+\times\R_x
\end{equation}
with
\begin{equation}\label{Vgeneral}
\begin{cases*}
0 < V\in\mathcal{C}^\infty(\R^+)\\
V'<0\quad \mathrm{and} \quad V\eqsim e^{-\lambda t}, ~~\forall t>t_{large}\ge0 \quad \mathrm{with} \quad \lambda>0.
\end{cases*}\end{equation}

In view of Lemma \ref{lem:properties_V_ell}, $V_\ell(t)$ satisfies \eqref{Vgeneral} with $\lambda = -2\kappa_-$ and $V_\ell(-t)$ satisfies \eqref{Vgeneral} with $\lambda = 2\kappa_+$. Hence, it is sufficient to study Equation \eqref{waveAnalytic} under the conditions \eqref{Vgeneral} with a general $\lambda>0$ while restricting to $t\ge0$.

\subsection{A Toy Model}\label{sec:toymodel}

In the following two sections, i.e., sections \ref{sec:toymodel} and \ref{sec:gen_case}, the variable $\omega$ will designate the spatial frequency, i.e., the Fourier dual variable of $x$. Starting from section \ref{sec:waveop} and on, we go back to using $\omega$ as the angular variable on the sphere $\mathcal{S}^2$.

We first study the special case where $V(t)=e^{-\lambda t}$, $\forall t\in\R$, which serves as a toy-model illustrating the methodology. For this potential $V$, the following proposition shows that for solutions of Equation \eqref{waveAnalytic} having unit energy at $t=0$, the energy at $t\to +\infty$ can be arbitrarily small.

%
\begin{prop1}\label{prop:Vexpo}
Consider Equation \eqref{waveAnalytic} with $V(t)=e^{-\lambda t}$ and the energy associated to it
\begin{equation}\label{enegryEu}
E[u](t)=\int_{\R_x} (\dl_t u)^2 + (\dl_x u)^2 + V u^2  \d x. 
\end{equation}
There exists a sequence $(u_n)_n$ of solutions to \eqref{waveAnalytic} such that $E[u_n](0)=1$ and $u_n(0,x)\in\mathcal{C}_c^\infty(\R_x)$ for all $n$, and
\begin{equation}
\lim\limits_{n\rightarrow\infty}\lim\limits_{t\rightarrow +\infty} E[u_n](t)=0.
\end{equation}
\end{prop1}
\begin{proof}
	Since we are going to give a detailed proof for the statement under the general conditions \eqref{Vgeneral}, we only include  the essential details here.
	
	Once we find the initial data corresponding to the claimed sequence, the result then follows from the well-posedness of the equation using the standard theory for hyperbolic PDEs. To do so, we first apply a Fourier transformation in $x$ to \eqref{waveAnalytic} to obtain an ODE in $t$ with frequency $\omega$:
	\begin{equation}\label{waveeqinwwithVexp}
	\hat{u}_\omega''(t)+(\omega^2 + e^{-\lambda t})\hat{u}_\omega(t)=0,
	\end{equation}
	where
	\begin{equation}
	\hat{u}_\omega(t):=\frac{1}{\sqrt{2\pi}}\int_{\R} u(t,x)e^{i\omega x}\d x .
	\end{equation}
	This equation can be solved explicitly using Bessel's functions $J_\alpha(z)$ that are defined via the Gamma function by
	\begin{equation}\label{besselfunctionJ}
	J_\alpha(z)=\sum_{m=0}^{\infty}\frac{(-1)^m}{m!\Gamma(m+\alpha+1)}\left(\frac{z}{2}\right)^{2m+\alpha}.
	\end{equation}
	Namely, a solution to \eqref{waveeqinwwithVexp} with initial data $(u(0),u'(0))=(b,c)$ has the form\footnote{$\Re(z)$ and $\Im(z)$ are respectively the real and the imaginary parts of a complex number $z$.}
	\begin{equation}\label{solutionformbessel}
		\hat{u}_\omega(t)=2\frac{\Im\left[J_W(\Lambda e^{\frac{-t}{\Lambda}})(bK_b+cK_c)\right]}{\Im\left(\overline{K_b} K_c\right)}
	\end{equation}
	where $\Lambda=\frac{2}{\lambda}$, $W=i\omega\Lambda$, $K_b(\omega)=J_{-1-W}(\Lambda)-J_{1-W}(\Lambda)$, and $K_c(\omega)=2J_{-W}(\Lambda)$. A straightforward calculation shows that if we choose $(b,c)=(-\Re(K_c(\omega)),\Re(K_b(\omega)))$ then the energy density $E_\omega[\hat{u}_\omega](t)=|\hat{u}'_\omega(t)|^2 + (\omega^2+e^{-\lambda t})|\hat{u}_\omega(t)|^2$ tends, uniformly in $\omega$, as $t\rightarrow+\infty$ to
$\displaystyle
		\frac{4\omega^2}{\vert\Gamma(W+1)\vert^2}
$, which in turn goes to zero as $\omega\to 0$.

To finish the proof, we take any smooth compactly supported function $\varphi$ on $\R$, say with $\Vert \varphi\Vert_{L^2}=1$, and we note that $-\Re(K_c(0))=-K_c(0)=2J_0(\Lambda)$ and $\Re(K_b(0))=K_b(0)=2J_{1}(\Lambda)$. It is not hard then to show that the sequence $$\psi_n(x):=\frac{2\varphi(\frac{x}{n})}{\sqrt{n}}(J_0(\Lambda),J_1(\Lambda))$$ of initial data gives rise after a normalization to the sequence of solutions we seek in the statement.
\end{proof}

One can easily see from the above proof that for $\omega=0$ the energy density of the solution to \eqref{waveeqinwwithVexp} tends to zero as $t$ tends to infinity when the initial data are $(J_0(\Lambda),J_1(\Lambda))$. Through the inverse Fourier transformation, $w=0$ solutions correspond to constants on the initial surface $t=0$, and although non-zero constants do not have finite energy at $t=0$ due to the $L^2(\R_x)$ component, they can be approximated by smooth compactly supported functions.  Thus, for small frequencies $\omega$, \eqref{waveeqinwwithVexp} can be considered as a perturbation of $\hat{u}''+e^{-\lambda t}\hat{u}=0$, and its solutions can be seen as approximating the solutions of the non-perturbed equation, which, up to a Fourier transformation, was the idea behind $\psi_n$ above. 

Alternatively, since we are interested in the asymptotic behavior of the energy when $V$ becomes small for large times, one can start from $u''(t)+w^2 u(t)=0$ and study the solutions of  
\begin{equation}\label{waveeqinwwithgeneralV}
u''(t)+(w^2+V(t)){u}(t)=0
\end{equation}
as perturbations. 
This is the approach that we follow in Section \ref{sec:gen_case}.

\subsection{General Case}
\label{sec:gen_case}

To see the potential $V(t)$ as a perturbation, we will construct our sequence of solutions to \eqref{waveeqinwwithgeneralV} by imposing data on surfaces $t=t_0$ with $t_0$ large enough. For simplicity, we assume $u$ to be real. Let us first note that the energy density 
\begin{equation}\label{energydensityVgeneral}
E_\omega[u](t)=(u'(t))^2 + (\omega^2+V(t))(u(t))^2
\end{equation}
is eventually a decreasing function of $t$. Indeed,
\begin{equation}\label{derivateofEomega}
E_\omega[u]'(t)=2u''u'+2(w^2+V)uu'+V'u^2=V'u^2
\end{equation}
and $V'(t)<0$ after a fixed $t_{large}$ by the assumption in \eqref{Vgeneral}. 

The following estimate will be useful for our calculations.
\begin{lem1}
	For $u$ a solution of \eqref{waveeqinwwithgeneralV} and $t>t_0>t_{large}$, we have
	\begin{equation}\label{linearboundonu}
	\vert u(t) \vert \lesssim \big((|\omega|+2)\vert u(t_0)\vert + \vert u'(t_0) \vert\big)(1+t-t_0). 
	\end{equation}
\end{lem1}
\begin{proof}
	We calculate,
	\begin{align}
	(u(t)^2)''&=2u'(t)^2+2u(t)u''(t)\\
	&=2u'(t)^2-2(\omega^2+V(t))u(t)^2\\
	&\le 2 E_\omega[u](t)\\
	&\le 2 E_\omega[u](t_0) \qquad \forall t>t_0>t_{large}.
	\end{align}
	Integrating twice both sides from $t_0$ to $t$, we get
	 \begin{align}
	 u(t)^2 &\le u(t_0)^2 + 2 u(t_0)u'(t_0)(t-t_0)+E_\omega[u](t_0)(t-t_0)^2 \\
	 & = u(t_0)^2 + 2 u(t_0)u'(t_0)(t-t_0)+\left(u'(t_0)^2 + (\omega^2+V(t_0))u(t_0)^2\right)(t-t_0)^2 \\
	 &\le 2\left(u(t_0)^2 + u'(t_0)^2(t-t_0)^2\right)+ \left(\omega^2+V(t_0))u(t_0)^2\right)(t-t_0)^2\\
	 &\le 2\left(u'(t_0)^2 + (1+\omega^2+V(t_0))u(t_0)^2\right)(1+t-t_0)^2.
	 \end{align}
	 Increasing $t_{large}$ if necessary, we can assume using \eqref{Vgeneral}  that $V(t_0)\le 1$ which then gives the estimate \eqref{linearboundonu}.
\end{proof}

We now generalize Proposition \ref{prop:Vexpo} to $V$ submitted to \eqref{Vgeneral} only. We first prove the result as a statement in the frequency domain.

\begin{lem1}\label{lem:Energy-w-toZeroVgeneral}
Let $n\geq 1$, $\omega\in [-\frac1n,\frac1n]$ and consider Equation \eqref{waveeqinwwithgeneralV} with $V$ satisfying \eqref{Vgeneral}. If $u_n$ is a solution to \eqref{waveeqinwwithgeneralV} such that $u(t_0(n))\ne 0$  and $u_n'(t_0(n))=0$ with $t_0(n):=\frac{\ln (n)}{\lambda}$, then for $n$ large enough, we have
 \begin{equation}
\lim\limits_{t\rightarrow +\infty}\frac{E_\omega[u_n](t)}{E_\omega[u_n](t_{0}(n))}\lesssim \frac{1}{n},
 \end{equation}
where the limit is uniform in $\omega\in [-\frac1n,\frac1n]$.
\end{lem1}

\begin{proof}
	As hinted at in the beginning of the section, we start with $t_0\ge t_{large}$ and a solution for     
	\begin{equation}\label{waveequinwONLY}
	\tilde{u}''(t)+w^2 \tilde{u}(t)=0
	\end{equation}
	with $\tilde{u}(t_0)=1$ and $\tilde{u}'(t_0)=0$, i.e., $\tilde{u}(t):=\cos(w(t-t_0))$. Next, let $u(t)$ be a solution of \eqref{waveeqinwwithgeneralV} such that $u(t_0)=1$ and $u'(t_0)=0$, and consider $v(t)=u(t)-\tilde{u}(t)$. Then, $v(t_0)=v'(t_0)=0$ and $v$ satisfies an inhomogeneous version of Equation \eqref{waveequinwONLY}:
	\begin{equation}\label{waveeqforvNonHomogeneous}
	v''(t)+\omega^2v(t)=h(t),
	\end{equation}
	where $h(t)=-V(t)u(t)$.
	The solution of \eqref{waveeqforvNonHomogeneous} with zero initial conditions is
	\begin{equation}\label{integralformofv-Volterra}
	v(t)=\int_{t_0}^{t} \frac{h(s)\sin(\omega(t-s))}{\omega} \d s.
	\end{equation}
	Note that this is just the Volterra integral equation for $u$. The idea is as in the proof of Proposition \ref{prop:Vexpo}: to take the limit of the energy density as $\omega$ goes to zero. From \eqref{integralformofv-Volterra} we immediately have for $t>t_0>t_{large}$,
	\begin{align}
	\vert v(t) \vert &=\left\vert-\frac{1}{\omega} \int_{t_0}^{t} V(s)u(s)\sin(\omega(t-s)) \d s \right\vert \\
	&\lesssim \frac{e^{-\lambda t_0}}{|\omega|} \int_{t_0}^{t} e^{-\lambda(s-t_0)}\vert u(s)\vert \d s \\
	&\lesssim  \frac{e^{-\lambda t_0}{(|\omega| +2)}}{|\omega|} \int_{t_0}^{t} e^{-\lambda(s-t_0)} (1+s-t_0)\d s \\
	&\lesssim \frac{e^{-\lambda t_0}{(|\omega| +2)}}{|\omega|} \int_{0}^{\infty} e^{-\lambda s} (1+s)\d s = \frac{e^{-\lambda t_0}{(|\omega| +2)}}{|\omega|}\frac{\lambda+1}{\lambda^2}\lesssim \frac{e^{-\lambda t_0}{(|\omega| +2)}}{|\omega|}.
	\end{align}
	where we have used \eqref{Vgeneral} then \eqref{linearboundonu}. This is sufficient to control the $w^2 u(t)^2$ term in the energy density \eqref{energydensityVgeneral}. To control the term $V(t)u(t)^2$ we need a polynomial bound in $t$ with no poles in $\omega$. For this, we use the fact that $\displaystyle \left|\frac{\sin(x)}{x}\right|\le 1$:
	\begin{align}
	\vert v(t) \vert &=\left\vert \int_{t_0}^{t} -\frac{V(s)u(s)\sin(\omega(t-s))}{\omega(t-s)} (t-s)\d s \right\vert \\
	&\lesssim \int_{t_0}^{t} e^{-\lambda s}\vert u(s)\vert (t-s) \d s \\
	&\lesssim  {{(|\omega| +2)}}\int_{t_0}^{t} e^{-\lambda s} (t-s)(1+s-t_0)\d s \\
	&\lesssim (|\omega| +2)t \int_{0}^{\infty} e^{-\lambda s}(1+s)\d s \lesssim (|\omega| +2)t.
\end{align}
	Similarly, to control the $u'(t)^2$ term, we  have,
	\begin{align}
	\vert v'(t) \vert &=\left\vert -\frac{1}{\omega}V(t)u(t)\sin(w(t-t)) - \int_{t_0}^{t} V(s)u(s)\cos(\omega(t-s)) \d s \right\vert \\
	&\lesssim e^{-\lambda t_0} \int_{t_0}^{t} e^{-\lambda(s-t_0)}\vert u(s) \vert \d s \\
	&\lesssim  {e^{-\lambda t_0}{(|\omega| +2)}}\int_{t_0}^{t} e^{-\lambda(s-t_0)} (1+s-t_0)\d s \\
	&\lesssim {e^{-\lambda t_0}{(|\omega| +2)}}.
	\end{align}
	It follows then from $u(t)=\cos(w(t-t_0))+v(t)$ that
	\begin{equation}
	\vert u(t) \vert \lesssim 1 + \frac{e^{-\lambda t_0}{(|\omega| +2)}}{|\omega|} ,
    \qquad\vert u(t) \vert \lesssim 1 + (|\omega| +2)t
    \qquad\mathrm{and}    \qquad
    \vert u'(t) \vert \lesssim |\omega| +  e^{-\lambda t_0}(|\omega| +2).
	\end{equation}
	Thence,
	
    \begin{align}
        E_\omega[u](t)&= (u'(t))^2 + (\omega^2+V(t))(u(t))^2 \\
    &\lesssim (|\omega| +  e^{-\lambda t_0}(|\omega| +2))^2 + \omega^2\left(1 + \frac{e^{-\lambda t_0}{(|\omega| +2)}}{|\omega|}\right)^2+V(t)\big(1+(|\omega|+2)t\big)^2.
\label{eq:EstimateEuw}
    \end{align}

    Let $n\geq e^{\lambda t_{large}}$ and $\displaystyle\omega\in[-\delta,\delta]$ with $\delta=\frac{1}{n}$ . Now choose $t_0=t_0(n):=\frac{\ln(n)}{\lambda}\geq t_{large}$ so that $e^{-\lambda t_0}=\delta$. Let $u_n$ be a solution to Equation \eqref{waveeqinwwithgeneralV} with  $u_n(t_0(n))=1$ and $u_n'(t_0(n))=0$. Under these assumptions, we have  
    \begin{equation}
    E_\omega[u_n](t_0(n))\gtrsim 0 + (w^2 + e^{-\lambda t_0(n)})\cdot 1 \gtrsim \delta\,,
    \end{equation}
 using \eqref{Vgeneral}. Furthermore, using \eqref{eq:EstimateEuw}, we get
    \begin{equation}
    E_\omega[u_n](t)\lesssim \delta^2 + V(t)(1+(\delta+2)t)\xrightarrow[t\rightarrow +\infty]{} \delta^2.
    \end{equation}
    Therefore,
     \begin{equation}
\lim\limits_{t\rightarrow +\infty}\frac{E_\omega[u_n](t)}{E_\omega[u_n](t_0(n))}\lesssim  \delta.
    \end{equation}
Since $E_w[C u_n](t)=C^2 E_w[u_n](t)$ for $C$ independent of $t$, and since we are interested in the quotient $$\displaystyle\frac{E_\omega[u_n](t)}{E_\omega[u_n](t_{0}(n))},$$
we lost no generality in assuming that $u_n(t_0(n)) = 1$.
\end{proof}

Now we can show that Proposition \ref{prop:Vexpo} holds for a general $V$ under the conditions \eqref{Vgeneral}:

\begin{prop1}\label{prop:EtozeroSequence-Vgeneral}
Consider Equation \eqref{waveAnalytic} with $V$ satisfying conditions \eqref{Vgeneral} and the energy associated to it
\begin{equation}\label{enegryEu}
E[u](t)=\int_{\R_x} (\dl_t u)^2 + (\dl_x u)^2 + V u^2  \d x. 
\end{equation}
There exists a sequence $(u_n)_n$ of solutions to \eqref{waveAnalytic} such that $E[u_n](0)=1$ and $u_n(0,x)\in\mathcal{C}_c^\infty(\R_x)$ for all $n$, and
\begin{equation}
\lim\limits_{n\rightarrow\infty}\lim\limits_{t\rightarrow +\infty} E[u_n](t)=0.
\end{equation}
\end{prop1}
\begin{proof}
	Clearly, it suffices to show that there exists a sequence $(u_n)_n$ of solutions to \eqref{waveAnalytic} with $u_n(0,x)\in\mathcal{C}_c^\infty(\R_x)$ for all $n$, such that 
	\begin{equation}
		\lim\limits_{n\rightarrow\infty}\lim\limits_{t\rightarrow +\infty} \frac{E[u_n](t)}{E[u_n](0)}=0.
	\end{equation}
Moreover, this is equivalent to taking the initial conditions at any fixed $\tilde{t}\in\R^+$ (in particular, $\tilde{t}=t_{large}$) instead of $t=0$, because $E[u_n](\tilde{t})\eqsim_{\tilde{t}} E[u_n](0)$. Indeed, for $u$ a solution of \eqref{waveAnalytic}, an integration by parts gives
\begin{align}
	\vert  E[u]'(t)\vert&=\left| \int_{\R_x} 2\dl_t u\left( \dl_t^2 u -\dl_x^2 u +Vu\right)\d x +\int_{\R_x} V' u^2  \d x \right|=\left| \int_{\R_x} V' u^2  \d x \right|\le  \int_{\R_x} \left|\frac{V'}{V}\right| V u^2  \d x \le \left|\frac{V'}{V}\right| E[u](t),
\end{align}
and then by Grönwall's inequality, we get
\begin{equation}\label{energyequivfrom-0-to-t}
	\displaystyle \exp\left(-\int_{0}^{\tilde{t}} \left|\frac{V'(s)}{V(s)}\right| \d s\right)\cdot E[u](0)\;\le \; E[u](\tilde{t}) \; \le \;\exp\left(\int_{0}^{\tilde{t}} \left|\frac{V'(s)}{V(s)}\right| \d s\right)\cdot E[u](0).
\end{equation}

Let $\varphi\in\mathcal{C}_c^\infty(\R_x)$ be an even function\footnote{The assumption that $\varphi$ is even was added for convenience so that its Fourier transform becomes real, but this is not indispensable.} such that $\Vert \varphi\Vert_{L^2}=1$. Let $n\geq e^{\lambda t_{large}}$ and $t_0(n):=\frac{\ln(n)}{\lambda}\geq t_{large}$, and let $u_n$ be the solution to \eqref{waveAnalytic} given by:
\begin{equation}
	\begin{cases}
		u_n(t_0(n),x)=\frac{1}{n}\varphi(\frac{x}{n^2}) \\
		\dl_{t}u_n(t_0(n),x)=0 .
	\end{cases}
\end{equation}
The Fourier transform $\hat{u}_n$ of $u_n$ solves \eqref{waveeqinwwithgeneralV} and satisfies:
 \begin{equation}
 	\begin{cases}
 		\hat{u}_n(t_0(n),\omega)=n\hat{\varphi}({\omega}{n^2}) \\
 		\dl_{t}\hat{u}_n(t_0(n),\omega)=0 ,
 	\end{cases}
 \end{equation}
and 
\begin{equation}
	E[u_n](t)=\int_{\R_\omega} E_\omega[\hat{u}_n](t)\d \omega=\int_{|\omega|>\frac{1}{n}} E_\omega[\hat{u}_n](t)\d \omega + \int_{|\omega|<\frac{1}{n}} E_\omega[\hat{u}_n](t)\d \omega =: I_>(t)+I_<(t).
\end{equation}
Recall from \eqref{derivateofEomega} that for $t>t_0(n)$, we have $E_\omega[u_n](t)\le E_\omega[u_n](t_0(n))$ for all $\omega\in\R$, and hence,
\begin{equation}
		E[u_n](t)\le I_>(t_0(n)) + I_<(t).
\end{equation}
Therefore, for $t>t_0(n)>t_{large}$, we have
\begin{equation}\label{energyfrac-t/tlarge}
	\frac{E[u_n](t)}{E[u_n](t_{large})}\le \frac{E[u_n](t)}{E[u_n](t_{0}(n))}\le  \frac{ I_>(t_0(n)) + I_<(t)}{ I_>(t_0(n)) + I_<(t_0(n))}
	=\displaystyle\frac{\displaystyle\frac{ I_>(t_0(n))}{I_<(t_0(n))} + \frac{I_<(t)}{I_<(t_0(n))}}{\displaystyle \frac{I_>(t_0(n))}{I_<(t_0(n))} + 1}.
\end{equation} 
From Lemma \ref{lem:Energy-w-toZeroVgeneral}, we have
\begin{equation}
\lim\limits_{t\rightarrow +\infty}	\frac{I_<(t)}{I_<(t_0(n))}=\frac{\displaystyle \int_{|\omega|<\frac{1}{n}}  \lim\limits_{t\rightarrow +\infty} \frac{ E_\omega[\hat{u}_n](t)}{E_\omega[\hat{u}_n](t_0(n))} E_\omega[\hat{u}_n](t_0(n))\d \omega}{I_<(t_0(n))}\lesssim \frac{1}{n}	\frac{I_<(t_0(n))}{I_<(t_0(n))}= \frac{1}{n},
\end{equation}
and thus,
\begin{equation}
	\lim\limits_{n\rightarrow\infty}\lim\limits_{t\rightarrow +\infty}	\frac{I_<(t)}{I_<(t_0(n))}=0.
\end{equation}
If we also show that 
\begin{equation}\label{limitfractiontozero}
	\lim\limits_{n\rightarrow\infty}	\frac{I_>(t_0(n))}{I_<(t_0(n))}=0,
\end{equation}
the proposition will be proved. 
Indeed, 
\begin{align}
	\frac{I_>(t_0(n))}{I_<(t_0(n))}&=\frac{\displaystyle\int_{|\omega|>n} \left({\omega^2}+n^4 V(t_0(n))\right)\hat{\varphi}(\omega)^2 \d \omega}{\displaystyle\int_{|\omega|<n} \left({\omega^2}{}+n^4V(t_0(n))\right)\hat{\varphi}(\omega)^2 \d \omega} \\
&=\frac{\displaystyle\int_{|\omega|>n} {\omega^2}\hat{\varphi}(\omega)^2\d \omega +n^4 V(t_0(n))\int_{|\omega|>n}\hat{\varphi}(\omega)^2 \d \omega}{\displaystyle\int_{|\omega|<n} {\omega^2}\hat{\varphi}(\omega)^2 \d \omega +n^4 V(t_0(n))\int_{|\omega|<n}\hat{\varphi}(\omega)^2 \d \omega}\\
	&\le \frac{\displaystyle\int_{|\omega|>n} {\omega^2}\hat{\varphi}(\omega)^2 \d \omega}{\displaystyle\int_{|\omega|<n} {\omega^2}\hat{\varphi}(\omega)^2\d \omega} + \frac{\displaystyle\int_{|\omega|>n} \hat{\varphi}(\omega)^2 \d \omega}{\displaystyle\int_{|\omega|<n} \hat{\varphi}(\omega)^2\d \omega},\label{lastline}
\end{align}
where the inequality follows from the fact that 
\begin{equation*}
	\frac{a+b}{c+d}= \frac{a}{c+d} + \frac{b}{c+d}\le \frac{a}{c} +\frac{b}{d}
\end{equation*}
whenever $a\ge0 , b\ge0, c>0,$ and $d>0$. Since $\hat{\varphi}$ and $ \omega\hat{\varphi}\in L^2(\R_\omega)$, we have 
\begin{align}
	\lim\limits_{n\rightarrow\infty}\int_{|\omega|>n} {\omega^2}\hat{\varphi}(\omega)^2 \d \omega=0  \quad &; \quad 	\lim\limits_{n\rightarrow\infty}\int_{|\omega|>n}\hat{\varphi}(\omega)^2 \d \omega=0 ~; \\
	 	\lim\limits_{n\rightarrow\infty}\int_{|\omega|<n} {\omega^2}\hat{\varphi}(\omega)^2 \d \omega=\Vert \omega \varphi \Vert_{L^2(\R_\omega)} > 0  \quad &; \quad 	\lim\limits_{n\rightarrow\infty}\int_{|\omega|<n}\hat{\varphi}(\omega)^2 \d \omega=1 ~.
\end{align} 
Therefore, the last line in \eqref{lastline} goes to zero as $n$ goes to infinity, proving \eqref{limitfractiontozero}.
\end{proof}

The following corollary is a restatement of Proposition \ref{enegryEu} in terms of $\mathcal{E}[\phi]$.

\begin{cor1}
\label{cor:enegryEphiUnbounded}
There exists a sequence $(\phi_n^{\pm})_n$ of solutions to \eqref{eq:geometricWaveEquation} such that $\phi_n^{\pm}|_{\Sigma_0}\in \mathcal{C}_c^\infty(\Sigma_0)$ and $\mathcal{E}[\phi_n^{\pm}](0)=1$ for all $n$, and 
\begin{equation}
\lim\limits_{n\rightarrow\infty}\lim\limits_{t\to \pm\infty} \mathcal{E}[\phi_n^{\pm}](t)=0.
\end{equation}
\end{cor1}
\begin{proof}
We choose $\ell\ge \ell_0$  so that Lemmata \ref{lem:equivEphiEu} and \ref{lem:properties_V_ell} hold. Now apply Proposition \ref{enegryEu} on $V(t):=V_{\ell}(\pm t)$ to get sequences  $(u_n^{\pm})_n$. Let $$\tilde\phi^{\pm}_{n}(t,x,\theta,\varphi)=\frac{u_n^{\pm}(\pm t,x)}{r} Y_{0,\ell}(\theta,\varphi),$$ where $Y_{0,\ell}$ is the $(m=0,\ell)$-th spherical harmonic. Since $u_n^{\pm}(\pm t,x)$ satisfy \eqref{wavedecompeq}, it is easy to see that $\tilde\phi^{\pm}_{n}$ satisfy $\eqref{eq:geometricWaveEquation}$. Now from Lemma \ref{lem:equivEphiEu} it follows that $$\phi^{\pm}_{n}:=\frac{\tilde\phi^{\pm}_{n}}{\sqrt{\mathcal{E}[\tilde\phi_n^{\pm}](0)}}$$ have the properties required in the corollary.
\end{proof}

\subsection{Wave Operators}\label{sec:waveop}
In this section, we will first study the scattering of solutions to Equation \eqref{waveAnalytic}, and then apply the same arguments to analyze the analytic scattering of Equation \eqref{eq:geometricWaveEquation}. 

One purpose is to show that a result analogous to Theorem \ref{mainthm} holds in the settings of \eqref{Vgeneral}, which are in some sense more general than the black hole setting. For example, one can consider the potential of Equation \eqref{waveAnalytic} as a time dependent and exponentially decaying mass of some field. This section shows that there is no scattering theory for this setting in the usual sense. Note that many scattering results are stated and studied using the wave operators approach. See e.g., \cite{hafner_scattering_2004,hafner_sur_2003,hafner_scattering_2021,borthwick_scattering_2020, besset_scattering_2021,reed_methods_1972,parra_spectral_2018,bony_scattering_2005,gerard_scattering_1998,nicolas_conformal_2016}.

Here, we look at the inverse wave operator that plays the role of the trace operator in the dynamical approach for scattering, which is based on comparison with simplified dynamics.
\subsubsection{Full Dynamics}
Let $\mathbb{H}(t)$ be the Hilbert space of finite $E$-energy data, i.e., $H^1(\R)\times L^2(\R)$ equipped with the norm
\begin{equation}
	\Vert (u,v)\Vert_{\mathbb{H}(t)}=\int_{\R} v^2 + (\dl_{x} u )^2 +V(t)u^2 \d x.
\end{equation}
\begin{note}\label{Rem:localuniformequivofHs}
	Note that $\mathbb{H}(t)\simeq_{t,s}\mathbb{H}(s)$ \textit{locally} uniformly in $(t,s)$.
\end{note}

We now rewrite  Equation \eqref{waveAnalytic}  in first order formulation. Let \begin{equation}\label{Firstorderreduction}
	\tilde{U}(t):=\left(\begin{array}{c}
		u(t)\\\dl_{t} u(t)
	\end{array}\right) \quad \mathrm{and} \quad P(t):=\dl_{x}^2-V(t),
\end{equation} then $u$ is a solution of \eqref{waveAnalytic} if and only if
\begin{equation}\label{waveeqFirstOrder}
	\dl_{t}	\tilde{U}(t)=iH(t)	\tilde{U}(t), \quad \mathrm{where} \quad H(t):=-i\left(\begin{array}{cc}
		0 & 1\\ P(t) & 0
	\end{array}\right).
\end{equation}
Here $D(H(t))$, the domain of $H(t)$, can be taken to be any suitable dense subset of $\mathbb{H}(t)$, such as smooth compactly supported pairs of functions on $\R$. Moreover, $\Vert 	\tilde{U}(t) \Vert_{\mathbb{H}(t)}=E[u](t)$ in the notations of \eqref{Firstorderreduction}.

Since \eqref{waveeqFirstOrder} is well-posed in $\mathbb{H}(s)$, the propagator associated to it $\mathcal{U}(t,s)$, which takes data in $\mathbb{H}(s)$ and gives the solution to \eqref{waveeqFirstOrder} at time $t$, is a strongly continuous evolution system. The following properties are easy to show:
\begin{note}\label{Rem:propertiesofPropagator}
 The evolution system	$\mathcal{U}(t,s)$ satisfies:
\begin{enumerate}
	\item For any $(s,t)\in \R^+\times\R^+$, $\mathcal{U}(t,s)\in\mathcal{L}(\mathbb{H}(s);\mathbb{H}(t))$.
    \item For any $t_1, t_2,t_3 \in \R^+$, $\mathcal{U}(t_3,t_2) \mathcal{U}(t_2,t_1) = \mathcal{U}(t_3,t_1)$; and for any $t \in \R^+$, $\mathcal{U}(t,t) =\mathrm{Id}_{\mathbb{H}(t)}$.
    \item For any $s\in \R^+$, $t \mapsto \mathcal{U}(t,s)$ is strongly continuous on $\mathbb{H}(s)$ (see Remark \ref{Rem:localuniformequivofHs}).
    \item For any $U\in\mathbb{H}(s)$ such that $\mathcal{U}(t,s) U\in D(H(t))$,
    \begin{equation}\label{evolution system1}
     \frac{\d}{\d t} \mathcal{U}(t,s) U = H(t) \mathcal{U}(t,s) U
    \end{equation}
\end{enumerate}
Moreover, the set $\mathcal{C}_c^\infty(\R)\times\mathcal{C}_c^\infty(\R)$ is stable under $\mathcal{U}(t,s)$ and $H(t)$ for all $t$ and $s$ in  $\R^+$. 
\end{note}

\subsubsection{Comparison Dynamics}\label{sec:comparisonDyn}
Since we have $\lim_{t\rightarrow + \infty} V(t)=0$, we define the asymptotic operator $P_+=\dl_{x}^2$. We use it to introduce the simplified wave equation which in first order formulation reads
\begin{equation}\label{waveeqSimplifiedFirstOrder}
	\dl_{t}	\tilde{U}(t)=iH_+	\tilde{U}(t), \quad \mathrm{where} \quad H_+:=-i\left(\begin{array}{cc}
		0 & 1\\\ P_+ & 0
	\end{array}\right).
\end{equation}
We thus define the asymptotic energy space to be $\mathbb{H}^+:=\dot{H}^1(\R)\times L^2(\R)$ with its natural norm, which we denote by $\Vert \cdot \Vert_{\mathbb{H}^+}$. The natural domain of $H_+$ is $D(H_+)=\{U\in\mathbb{H}^+; H_+U\in \mathbb{H}^+\}$, and $(H_+,D(H_+))$ is a self-adjoint operator on $\mathbb{H}^+$. This  entails that the propagator of \eqref{waveeqSimplifiedFirstOrder} is given by the 1-parameter family $e^{itH_+}$ of  unitary operators on $\mathbb{H}^+$ by Stone's theorem. 

\subsubsection{Inverse Wave Operator}

We are now in place to state and prove the theorem constructing the inverse wave operator associated with the full dynamics of Equation \eqref{waveeqFirstOrder} and the free dynamics of \eqref{waveeqSimplifiedFirstOrder}.    

\begin{thm1}\label{thm:Inv-wave-op}
	The inverse wave operator $\Omega$, defined by the following strong limit with respect to the norm $\Vert \cdot \Vert_{\mathbb{H}^+}$
	\begin{equation}\label{stronglimit}
		\Omega U= \lim_{t\rightarrow + \infty} e^{-itH_+}\mathcal{U}(t,0)U\quad\quad\text{for \;$U\in\mathcal{C}_c^\infty(\R)\times\mathcal{C}_c^\infty(\R)$},
	\end{equation}
extends to a bounded linear transformation from $\mathbb{H}(0)$ to $\mathbb{H}^+$. 

However, the extended inverse wave operator $\Omega$ does not admit a bounded inverse. 
\end{thm1}
\begin{proof}
	We first need to prove that the limit in \eqref{stronglimit} exists. For this we use Cook's  method (see e.g., \cite[Section XI.3]{reed_methods_1972}) by showing that
	\begin{equation}\label{derivativeinL1}
	 \frac{\d}{\d t}(e^{-itH_+}\mathcal{U}(t,0)U) \in L^1(\R^+;\mathbb{H}^+),\quad\forall U\in\mathcal{C}_c^\infty(\R)\times\mathcal{C}_c^\infty(\R).
	\end{equation}
We calculate using Remark \ref{Rem:propertiesofPropagator} and the fact that the energy is decreasing after $t_{large}$,
\begin{align}
	\left\Vert \frac{\d}{\d t}(e^{-itH_+}\mathcal{U}(t,0)U) \right\Vert_{\mathbb{H}^+}^2&=\left\Vert i e^{-itH_+}(-H_+ + H(t))\mathcal{U}(t,0)U \right\Vert_{\mathbb{H}^+}^2 \\
	&=\left\Vert  \left(\begin{array}{cc}
		0 & 0\\ -V(t) & 0
	\end{array}\right) \mathcal{U}(t,0)U \right\Vert_{\mathbb{H}^+}^2\\	
	&\le V(t)\left\Vert\mathcal{U}(t,0)U \right\Vert_{\mathbb{H}(t)}^2\\
	&=V(t)\left\Vert\mathcal{U}(t,t_{large})\mathcal{U}(t_{large},0)U \right\Vert_{\mathbb{H}(t)}^2\\
	&\le V(t)\left\Vert\mathcal{U}(t_{large},0)U \right\Vert_{\mathbb{H}(t_{large})}^2
\end{align}
which goes exponentially to zero as $t\to +\infty$, showing \eqref{derivativeinL1}.

To show that $\Omega$ is bounded we again take $U\in\mathcal{C}_c^\infty(\R)\times\mathcal{C}_c^\infty(\R)$, then
\begin{align}
	\left\Vert \Omega U \right\Vert_{\mathbb{H}^+}&=\lim_{t\rightarrow + \infty}\left\Vert e^{-itH_+}\mathcal{U}(t,0)U \right\Vert_{\mathbb{H}^+}\\
	&=\lim_{t\rightarrow + \infty}\left\Vert \mathcal{U}(t,0)U \right\Vert_{\mathbb{H}^+}\\
	&\leq\lim_{t\rightarrow + \infty}\left\Vert \mathcal{U}(t,0)U \right\Vert_{\mathbb{H}(t)}\label{firstineq}\\
	&\le\left\Vert \mathcal{U}(t_{large},0)U \right\Vert_{\mathbb{H}(t_{large})}
	\le C_{t_{large}}\Vert U \Vert_{\mathbb{H}(0)},
\end{align}
where $C_{t_{large}}$ can be chosen to be the operator norm of $\mathcal{U}(t_{large},0)$ which in turn can be estimated from $E[u](t_{large})\eqsim_{t_{large}} E[u](0)$ as in \eqref{energyequivfrom-0-to-t}. Therefore, $\Omega$ extends as a bounded linear transformation from $\mathbb{H}(0)$ to $\mathbb{H}^+$. 

The unboundedness of the inverse of $\Omega$ can be obtained from Proposition \ref{prop:EtozeroSequence-Vgeneral} and the first inequality, \eqref{firstineq}, in the above calculation of $\left\Vert \Omega U \right\Vert_{\mathbb{H}^+}$.
\end{proof}

\subsubsection{Black Hole Case}

We now apply the same techniques as the previous paragraphs to analyze the analytic scattering of solutions to the wave equation $\square \phi=0$ on $\mathcal{M}$ in the $(t,x,\omega)$-coordinates, i.e., 
	\begin{equation}\label{wave-eq-Geom-in-Coord}
	\dl_t^2\phi-\dl_x^2\phi+\frac{2f}{r}\dl_t\phi +\frac{f}{r^2} \Delta_{{\mathcal{S}^2}}\phi=0. 
\end{equation}

Our Hilbert space on $\Sigma_t$ will be the same as in Section \ref{sec:FiniteEnergySpaces} with the norm given in \eqref{energynorm-t}, i.e. $\mathcal{H}(t)$, but we now see it as $$\left(\dot{H}^1(\R\times\mathcal{S}^2)\times L^2(\R\times\mathcal{S}^2),\Vert \cdot\Vert_{\mathcal{E}(t)}\right),$$
and a remark similar to Remark \ref{Rem:localuniformequivofHs} holds. We next reformulate \eqref{wave-eq-Geom-in-Coord} in first order: Set
\begin{equation}\label{FirstorderreductionBLACKHOLE}
	\tilde{\Phi}(t):=\left(\begin{array}{c}
		\phi(t)\\\dl_{t} \phi(t)
	\end{array}\right) \quad \mathrm{and} \quad P_B(t):=\dl_{x}^2-\frac{f}{r^2} \Delta_{{\mathcal{S}^2}},
\end{equation} then \eqref{wave-eq-Geom-in-Coord} is equivalent to
\begin{equation}\label{waveeqFirstOrderBLACKHOLE}
	\dl_{t}\tilde{\Phi}(t)=iH_B(t)\tilde{\Phi}(t), \quad \mathrm{where} \quad H_B(t):=-i\left(\begin{array}{cc}
		0 & 1\\ P_B(t) & \displaystyle -\frac{2f}{r}
	\end{array}\right).
\end{equation}
As before, we denote by $\mathcal{U}_B(t,s)$ the propagator associated with $H_B$. It satisfies  similar properties to those in Remark \ref{Rem:propertiesofPropagator}. Also, recall that $\Vert\mathcal{U}_B(t,0)\Phi\Vert_{\mathcal{E}(t)}^2=\mathcal{E}[\phi](t)$ where $\mathcal{U}_B(t,0)\Phi=(\phi(t),\dl_{t}\phi(t))$ with $\phi$ the  solution to \eqref{wave-eq-Geom-in-Coord} corresponding to some initial data $\Phi$.

The comparison dynamics will be as in Section \ref{sec:comparisonDyn}, with a slight difference in the Hilbert space. Namely,
\begin{equation}\label{waveeqSimplifiedFirstOrderBLACKHOLE}
	\dl_{t}\tilde{\Phi}(t)=iH_\infty\tilde{\Phi}(t), \quad \mathrm{where} \quad H_{\infty}:=-i\left(\begin{array}{cc}
		0 & 1\\\ \dl_x^2 & 0
	\end{array}\right).
\end{equation}
Here, $H_{\infty}$ is a linear operator on $\mathbb{H}_B^\infty:=\dot{H}^1(\R;L^2(\mathcal{S}^2))\times L^2(\R\times \mathcal{S}^2)$ with the norm $$\Vert (\phi_0,\phi_1) \Vert_{\mathbb{H}_B^\infty}:=\left(\hf\int_{\R\times\mathcal{S}^2} \phi_1^2+(\dl_x\phi_0)^2  \d x  \d^2\omega\right)^\hf.$$ 
Although it is more natural  in accordance with $\Vert \cdot\Vert_{\mathcal{E}(t)}$  to have a factor of $r^2_\pm$ in the $\Vert \cdot \Vert_{\mathbb{H}_B^\infty}$ norm (corresponding to the different limits of $r(t)$), we drop it in favor of simplicity since the function $r(t)$ is positive, smooth, and bounded on $\R_t$. This way, we have the same function space on future and past infinities.

Again, $H_\infty$ with its natural domain is a self-adjoint operator on $\mathbb{H}_B^\infty$, and the propagator $e^{itH_\infty}$ is unitary.  

\begin{thm1}\label{thm:inverse-wave-blackhole}
	The inverse wave operators $\Omega_B^\pm$, defined by the following strong limits with respect to the norm $\Vert \cdot \Vert_{\mathbb{H}_B^\infty}$
	\begin{equation}\label{stronglimitBH}
		\Omega_B^\pm \Phi= \lim_{t\rightarrow \pm \infty} e^{-itH_\infty}\mathcal{U}_B(t,0)\Phi\quad\quad\text{for \;$\Phi\in\mathcal{C}_c^\infty(\R\times\mathcal{S}^2)\times\mathcal{C}_c^\infty(\R\times\mathcal{S}^2)$,}
	\end{equation}
	extend to bounded linear transformations from $\mathbb{H}_B(0)$ to $\mathbb{H}_B^\infty$. 
	
	However, the extended inverse wave operators $\Omega_B^\pm$ do not admit  bounded inverses. 
\end{thm1}
\begin{proof}
	The proof is very similar to the proof of Theorem \ref{thm:Inv-wave-op}. Let $\Phi\in\mathcal{C}_c^\infty(\R\times\mathcal{S}^2)\times\mathcal{C}_c^\infty(\R\times\mathcal{S}^2)$, and let $\mathcal{U}_B(t,0)\Phi=(\phi(t),\dl_{t}\phi(t))$, i.e., $\phi$ is the corresponding solution to \eqref{wave-eq-Geom-in-Coord}. For Cook's method we have,
	\begin{align}
		\left\Vert \frac{\d}{\d t}(e^{-itH_\infty}\mathcal{U}_B(t,0)\Phi) \right\Vert_{\mathbb{H}_B^\infty}^2&=\left\Vert i e^{-itH_\infty}(-H_\infty + H_B(t))\mathcal{U}_B(t,0)\Phi \right\Vert_{\mathbb{H}_B^\infty}^2 \\
		&=\left\Vert  \displaystyle\left(\begin{array}{cc}
			0 & 0\\ \displaystyle-\frac{f}{r^2} \Delta_{\mathcal{S}^2} & \displaystyle-\frac{2f}{r}
		\end{array}\right) \mathcal{U}_B(t,0)\Phi \right\Vert_{\mathbb{H}_B^\infty}^2\\	
		&= \left\Vert \frac{f}{r^2} \Delta_{{\mathcal{S}^2}}\phi(t)+\frac{2f}{r}\dl_t\phi(t) \right\Vert_{L^2(\R\times\mathcal{S}^2)}^2\\
				&\lesssim f^2\left\Vert  \Delta_{{\mathcal{S}^2}}\phi(t) \right\Vert_{L^2(\R\times\mathcal{S}^2)}^2+ f^2\left\Vert r\dl_t\phi(t)\right\Vert_{L^2(\R\times\mathcal{S}^2)}^2\\
								&\lesssim - f^2 \int_{\R\times\mathcal{S}^2} \d \omega^2 \left(\nabla_{{\mathcal{S}^2}}\phi(t) , \nabla_{{\mathcal{S}^2}}\Delta_{{\mathcal{S}^2}}\phi(t)\right) \d x\d\omega + f^2\mathcal{E}[\phi](t)\\
&\le \resizebox{0.7\textwidth}{!}{$\displaystyle  -f \int_{\R\times\mathcal{S}^2} (-f)|\nabla_{{\mathcal{S}^2}}\phi(t)|^2 \d x\d\omega -f \int_{\R\times\mathcal{S}^2} (-f)|\nabla_{{\mathcal{S}^2}}\Delta_{{\mathcal{S}^2}}\phi(t)|^2 \d x\d\omega + f^2\mathcal{E}[\phi](t)$}\\
		&\lesssim (-f+f^2)\mathcal{E}[\phi](t) -f\mathcal{E}[\Delta_{{\mathcal{S}^2}}\phi](t),	\end{align}
	and integrability follows from the energy estimates \eqref{energy-estimates-t-positive} and \eqref{energy-estimates-t-negative}, and the fact that $f$ decays exponentially as $t\to\pm\infty$.
	
The boundedness of  $\Omega_B^\pm$ also immediately follows from \eqref{energy-estimates-t-positive} and \eqref{energy-estimates-t-negative}:
\begin{align}
	\left\Vert \Omega_B^\pm \Phi \right\Vert_{\mathbb{H}_B^\infty}&=\lim_{t\rightarrow \pm \infty}\left\Vert e^{-itH_\infty}\mathcal{U}_B(t,0)\Phi \right\Vert_{\mathbb{H}_B^\infty}\\
	&=\lim_{t\rightarrow \pm \infty}\left\Vert \mathcal{U}_B(t,0)\Phi \right\Vert_{\mathbb{H}_B^\infty}\\
	&\lesssim\lim_{t\rightarrow \pm \infty}\left\Vert \mathcal{U}_B(t,0)\Phi \right\Vert_{\mathcal{E}(t)}\label{secondineq}\\
	&\lesssim\left\Vert \Phi \right\Vert_{\mathcal{E}(0)}.
\end{align}	

Again, the unboundedness of the inverses of $\Omega_B^\pm$ is a consequence of Corollary \ref{cor:enegryEphiUnbounded} and inequality \eqref{secondineq}.
\end{proof}

\begin{note}
	The inverse wave operators can be explicitly identified with the trace operators using the flows of congruences of null geodesics. This equivalence has been previously established in several situations,  \cite{mason_conformal_2004,hafner_scattering_2021,nicolas_conformal_2016}, including the wave equation. The current situation is not so different, and we therefore skip the details.  
\end{note}

\printbibliography[heading=bibintoc] 
\end{document}